# Thin Film Growth and Device Fabrication of Iron-Based Superconductors

Hidenori Hiramatsu [1,*], Takayoshi Katase [1], Toshio Kamiya [1], and Hideo Hosono [1, 2]

1: Materials and Structures Laboratory, Tokyo Institute of Technology, Mailbox R3-1, 4259 Nagatsuta-cho, Midori-ku, Yokohama 226-8503, Japan

2: Frontier Research Center, Tokyo Institute of Technology, Mailbox S2-13, S2-6F East, 4259 Nagatsuta-cho, Midori-ku, Yokohama 226-8503, Japan

ABSTRACT

Iron-based superconductors have received much attention as a new family of high-temperature superconductors owing to their unique properties and distinct differences from cuprates and conventional superconductors. This paper reviews progress in thin film research on iron-based superconductors since their discovery for each of five material systems with an emphasis on growth, physical properties, device fabrication, and relevant bulk material properties.

(*) E-mail address: h-hirama@lucid.msl.titech.ac.jp

## 1. Introduction

The bulk superconductivity of LaFePO at a critical temperature ($T_c$) of ~4 K was discovered in 2006 [1] during a systematic study on electromagnetic functionalities in transition-metal oxypnictides with a ZrCuSiAs type structure (called '1111 compounds') and related pnictides [2–5]. Subsequently, unique properties were found in the La*TMPn*O (*TM* = 3*d* transition metal, *Pn* = pnictogen) system and in the phosphide; LaCo*Pn*O are ferromagnetic metals [6], while LaZn*Pn*O and LaMnPO are nonmagnetic and antiferromagnetic semiconductors, respectively [7,8]. Superconductivity was observed when $TM$ = $Fe^{2+}$ with a $3d^6$ electronic configuration or $Ni^{2+}$ with a $3d^8$ configuration, LaNi*Pn*O [9,10], F-doped LaFeAsO [11] with a ZrCuSiAs type structure, and $BaNi_2P_2$ [12] with a $ThCr_2Si_2$ type structure (called '122 compounds'). These materials are Pauli paramagnetic metals and exhibit bulk superconducting transitions at ≥ 2 K.

In particular, the superconductivity of F-doped LaFeAsO with $T_c$ = 26 K [11] had a strong impact on the superconductor research community. Applying an external pressure of 4 GPa enhanced its $T_c$ to 43 K [13], but replacing La with other small rare-earth ions (Ce–Sm) further increased $T_c$ to 55 K [14–20]. Carrier doping of parent compounds with antiferromagnetic ordering is required for such a high $T_c$ to emerge. Currently, the highest $T_c$ is 56 K, observed in Th-doped GdFeAsO [21].

Furthermore, three types of Fe-based parent compounds and their superconductivities were successively reported: $AE$Fe$_2$As$_2$ ($AE$ = alkaline earth such as Ba, Sr, or Ca) [22] with a ThCr$_2$Si$_2$ type structure, $A$FeAs ($A$ = alkali metal such as Na or Li, called '111 compounds') [23] with a CeFeSi type structure, and FeSe ('11 compounds') [24] with an anti-PbO type structure. Each compound has a layered tetragonal lattice structure constructed from the layers of edge-shared FeAs$_4$/FeSe$_4$ tetrahedra. In addition, a layered oxypnictide [25] with thick perovskite-like oxide-block layers joined the Fe-based superconductor family in 2009. Several derivatives of similar oxypnictide superconductors [26–34], which possess thick oxide layers, have been reported, and the maximum $T_c$ among these compounds is 47 K in (Fe$_2$As$_2$)(Ca$_4$(Mg$_{0.25}$Ti$_{0.75}$)$_3$O$_{7.5}$) [32]. This $T_c$ is close to the maximum value for the 1111 compounds.

The five parent phases mentioned above provide not only diverse relationships between superconductivity and magnetism [35–42], but also a large platform for research on superconducting thin films and devices. Superior superconducting properties such as a high $T_c \geq 55$ K [19–21] and a large upper critical magnetic field > 100 T [43] appear to be more appropriate for wire/tape applications under a high magnetic field. Additionally, Fe-based superconductors significantly differ from cuprates; the parent compounds of Fe-based superconductors are metals in normal states and exhibit superconductivity in a tetragonal symmetry, whereas those of cuprates are Mott insulators and exhibit superconductivity in an orthorhombic symmetry. Fe-based superconductors have lower anisotropy in the superconducting properties than cuprates (the 122 compounds have particularly low anisotropy among Fe-based superconductors [44]). Moreover, the intrinsic nature of superconductivity in Fe-based superconductors differs from that in cuprates; the former has a multi pocket structure on a Fermi surface and a higher symmetric order parameter ($s$-wave), but the latter has $d$-wave pairing [45–47]. Regardless, cuprates and Fe-based superconductors have several features in common, including a layered structure, superconductivity induced by carrier doping, and the presence of competing antiferromagnetic orders.

In July 2008, a novel doping method involving the partial substitution of Fe ions with Co was reported to induce superconductivity in 1111 [48–50] and 122 [51,52] compounds. Hereafter, this method is referred to as 'direct doping' because the carriers are injected directly to the conducting FeAs layer. The effectiveness of direct doping is unique to Fe-based superconductors because superconductivity is, in general, severely degraded by disturbances, caused by direct doping, to substructures controlling the Fermi surface. Therefore, this method has attracted much attention from the perspectives of its mechanism and a new doping mode. Moreover, this finding has also contributed to the fabrication of thin films and devices using Fe-based superconductors because Co possesses a low vapor pressure and is incorporated in thin films more easily than other dopants with high vapor pressures such as F for 1111 phases [11,14] and K for 122 phases [22]. Consequently, direct doping has led to the realization of high-quality and high critical current density ($J_c$) thin films of Co-doped Ba122.

In this article, we review the progress and current status of the growth and performance of thin films as well as the device fabrication of Fe-based superconductors. Similar to bulk synthesis, the technical difficulties in thin film fabrication strongly depend on the material system. Thus, this review article is divided into the following categories: §2. 1111 thin films, §3. 122 thin films, §4. 11 thin films, §5. superconducting thin film devices, and §6. brief summary of the latest results. To date, a paper has yet to report 111 thin films or oxypnictides thin films with thick oxide-block layers probably due to technical difficulties; i.e., 111 compounds have an alkali metal as the main component, and the film growth of the oxypnictides is difficult for the same reason as that of 1111 films.

## 2. 1111 Thin Films

Hiramatsu and co-workers have successfully grown epitaxial films of a variety of materials with the ZrCuSiAs type structure such as LaCu*Ch*O (*Ch* = chalcogen) [53] and LaZn*Pn*O [54] by reactive solid-phase epitaxy (R-SPE) [55], in which laser ablation with a KrF excimer laser (wavelength, $\lambda$ = 248 nm) is used as the deposition technique. On the other hand, LaMn*Pn*O [56] was fabricated by direct deposition using conventional laser ablation with an ArF excimer laser ($\lambda$ = 193 nm) as the excitation source. Additionally, thin films of other (oxy/fluoride)chalcogenides with the ZrCuSiAs type structure were also grown directly by conventional laser ablation using a KrF excimer laser [57–60].

Therefore, Hiramatsu et al. [61] attempted to grow epitaxial films of LaFeAsO (La1111) using several techniques, including conventional laser ablation, post-deposition thermal annealing in evacuated ampoules, and R-SPE using Fe metal as a sacrificial layer, which are all effective for growing epitaxial films of the other ZrCuSiAs type compounds as explained above. However, not even a polycrystalline film of the target La1111 phase was obtained, even when a single-crystal substrate was used. This difficulty in obtaining the La1111 phase sharply contrasts with the cases of the other ZrCuSiAs type compounds such as LaCu*Ch*O reported previously. Hence, they reoptimized the preparation conditions of bulk La1111 samples and obtained a pure-phase laser ablation target free from wide-gap impurity phases such as $La_2O_3$ and LaOF because such wide-gap phases are easily incorporated in the resulting films if they are included in the laser ablation target. Next, they changed the excitation source of laser ablation from an ArF excimer laser to the second harmonics ($\lambda$ = 532 nm) of a neodymium-doped yttrium aluminum garnet (Nd:YAG) laser, because they speculated that the active oxygen species generated by ultraviolet light would negatively affect film growth [Fig. 1(a)]. These improvements along with further optimization of the substrate temperature to ~780 °C under a vacuum of ~$10^{-5}$ Pa led to the successful growth of epitaxial La1111 thin films on three types of (001)-oriented cubic single-crystal substrates with different lattice mismatches: MgO (in-plane lattice mismatch +4%), $MgAl_2O_4$ (+0.1%), and (La, Sr)(Al, Ta)$O_3$ (LSAT, –4%). This was the first demonstration of La1111 epitaxial films. Unfortunately, none of the La1111 epitaxial films exhibited a superconducting transition, at least down to 2 K, as shown in Fig. 1(b). The resistivity–temperature ($\rho$–$T$) curves of the thin films were similar to those of undoped bulk samples; i.e., an anomaly of resistivity was observed at ~150 K, which originates from the structural [62] and magnetic [63] transitions. This result suggests that fluorine dopants were not incorporated in the films even though the laser ablation target had a high fluorine concentration of 10 at.%.

Shortly after that, Backen et al. [64] of Holzapfel's group reported the fabrication of biaxially textured La1111 thin films on MgO and $LaAlO_3$ (LAO) (001) substrates, but the intensities of the out-of-plane 110 and 112 x-ray diffraction (XRD) peaks were weak. They employed a combined method of laser ablation using a KrF excimer laser at room temperature for film deposition with subsequent post-deposition thermal annealing in an evacuated silica-glass ampoule for crystallization. By optimizing the annealing time, they obtained a maximum $T_c^{onset}$ of 11 K (Fig. 2). Although zero resistance was not observed, superconductivity was confirmed by applying external magnetic fields. The fluorine concentration of the ablation target was as large as 25 at.%, suggesting that it is difficult to incorporate a sufficient fluorine concentration to induce superconductivity in 1111 films. A similar difficulty in fluorine incorporation was also reported in the case of molecular beam epitaxy (MBE) by Kawaguchi et al. [65] of Ikuta's group. They successfully obtained epitaxial films of NdFeAsO (Nd1111) on GaAs (001) substrates by choosing suitable sources and optimizing their fluxes. Although the process window was very narrow, the reproducibility was good. We consider that this is an advantage of MBE over laser ablation. The phase purity of the films obtained under optimum conditions was very high, but superconductivity was not observed at ≥ 2 K (Fig. 3). The $\rho$–$T$ behavior was very similar to that of thin films fabricated by laser ablation [61], suggesting that fluorine was not effectively incorporated in their MBE films.

During the initial research on 1111 film growth, the formation of the 1111 phase and incorporation of the fluorine dopant in thin films were the largest obstacles to obtaining superconducting films. However, Holzapfel and Ikuta's groups have overcome these issues and successfully doped F in their 1111 films. Haindl et al. [66] achieved La1111 films, which exhibited clear superconducting transitions at $T_c^{onset}$ = 28 K and $T_c^{zero}$ = ~20 K, by reducing the oxygen partial pressure during post-deposition thermal annealing (Fig. 4). This was the first report of a 1111 thin film that clearly had zero resistance. Its self-field $J_c$ at a low temperature is < 2 kA/cm$^2$, implying that the film has granular microstructures and exhibits weak-link behavior. Additionally, Kidszun et al. [67] obtained epitaxially grown La1111 films by further improving the thermal annealing atmosphere. Their research has contributed to a recent report on thin film fabrication of Gd1111, which is expected to realize a higher $T_c$ (because $T_c$ of bulk Gd1111 is 54–56 K [18,21].) [68]. Such improvements have also led to a recent report on the higher $J_c$ of La1111 films (~0.1 MA/cm$^2$ at 4.2 K) as well as a discussion on the angular dependence of $J_c$ [69]. Using high-quality La1111 epitaxial films, they demonstrated that $J_c$ can be scaled using the Blatter approach [70] with temperature-dependent anisotropy and concluded that F-doped La1111 has two major weakly coupled bands.

In the case of MBE growth, Kawaguchi et al. [71] found that the growth time ($t_g$) strongly affects the superconducting properties of Nd1111 films. They successfully obtained single-phase epitaxial Nd1111 films when $t_g$ ≤ 3 h, but the films did not exhibit superconductivity, similar to in their previous study [65]. On the other hand, when $t_g$ ≥ 3 h, the Nd1111 phase was still the major one, but some impurities (Fe$_2$O$_3$, FeAs, and NdOF) appeared. Upon further increasing $t_g$, the formation of a NdOF impurity became marked in the samples with $t_g$ = 5 and 6 h. However, their physical properties drastically changed between $t_g$ = 4 and $t_g$ ≥ 5 h (Fig. 5). Clear superconducting transitions were observed for the samples with $t_g$ = 5, 6 h, and $T_c$ of the $t_g$ = 6 h sample ($T_c^{onset}$ = 48 K and $T_c^{zero}$ = 42 K) was slightly higher than that of the $t_g$ = 5 h sample. This observation implies that the fluorine dopant can be effectively introduced when $t_g$ > 4 h. They confirmed that fluorine was detected only in the samples of $t_g \geq 5$ h and that a longer deposition resulted in the segregation of the NdOF impurity at the film surface, similar to in the report by Kidszun et al. [67]. Therefore, a general trend may be that an oxyfluoride impurity segregates easily at surfaces of 1111 films.

In this section, we summarized the current status of 1111 thin films. Only four research groups have successfully fabricated 1111 thin films (recently, Naito's group also reported Sm1111 superconducting films grown by MBE [72].), demonstrating the difficulties in film growth and fluorine doping. The origin of these difficulties and the growth mechanism remain unclear. $T_c$ reported for 1111 thin films is approaching that of bulk samples, and $T_c$ at 56 K, which is the highest $T_c$ reported to date for Fe-based superconductor thin films, should be reported in the very near future [73]. We suggest that MBE may be an effective method of obtaining 1111 thin films with a high $T_c$ because each element source and flux rate can be separately controlled. Regarding $J_c$, only Holzapfel's group has reported a high $J_c$ of approximately 0.1 MA/cm$^2$ at 4.2 K and scaling anisotropy [69]. The $J_c$ is comparable to those of single crystals [74–76] but slightly lower than those in a few reports [77,78]. The employment of magnetic and optical measurements and/or further consideration of the results obtained by other groups may lead to the development of more sophisticated growth technology for 1111 thin films. Future studies should clarify the physical properties of the 1111 phase and demonstrate superconducting devices accompanying a high $T_c$.

## 3. 122 Thin Films
### 3.1 Sr system

Shortly after the report on epitaxially grown La1111 [61], Hiramatsu et al. [79] tried to grow epitaxial films of Co-doped SrFe$_2$As$_2$ (Sr122:Co), whose bulk has a $T_c$ of ~20 K [52], using the Nd:YAG laser ablation system shown in Fig. 1(a). Because the vapor pressures of Co-related species are much lower than those of F (for 1111) and K (for 122) compounds, they speculated that Co doping (direct doping) should be easier than F/K doping in 1111/122 thin films. Additionally, they anticipated that its growth would be easier than a mixed-anion 1111 phase compound in which O and As coexist because the 122 compound has only one anion (As). Setting the substrate temperature to ~600 °C produced a strongly $c$-axis oriented Sr122:Co film, but in-plane orientation was not confirmed. On the other hand, at ~700 °C, epitaxial growth on an LSAT (001) single-crystal substrate was confirmed using the epitaxial relationship (001)[100] SrFe$_2$As$_2$ ∥ (001)[100] LSAT [Fig. 6(a)]. The $\rho$–$T$ curve of the Sr122:Co film grown at ~700 °C exhibited a clear superconducting transition at $T_c^{onset}$ = 20 K and $T_c^{zero}$ = 14 K ($\Delta T_c$ = 6 K) [Fig. 6(b)]. This was the first fabrication of a 122 superconducting film. This film contributed to clarifying that the upper critical magnetic field of an epitaxial film is ~47 T, and that the linear decrease in anisotropy is described well by a two-band model (Fig. 7) [80]. Evaluating the angular dependence of $J_c$ and the melting line (Fig. 8) of the Sr122:Co films [81] indicates that the typical self-field $J_c$ is as low as 10–20 kA/cm$^2$ at 4.2 K, confirming that the films have a granularity problem (weak link). In addition, two $J_c$ peaks at $H \parallel c$ and $H \perp c$ and a magnetic history effect were observed, implying that magnetic pinning, which may be due to the Fe metal impurity, is plausible.

Later, Choi et al. [82] successfully fabricated Sr122:Co thin films on LAO (001) substrates by laser ablation. They used As-rich ablation targets to compensate the As deficiency in the films and employed a KrF excimer laser as the excitation source. The obtained superconducting properties were $T_c^{onset}$ = 18.6 K, $T_c^{zero}$ = 16.4 K ($\Delta T_c$ = 2.2 K), and $J_c$ (5 K, 0.1 T) = 16 kA/cm$^2$. Such a low $J_c$ implies that their films also suffered from granularity and/or magnetic impurity problems. However, this report clarifies that, unlike 1111 films [61], a Nd:YAG laser is not necessary to grow 122 films; indeed a KrF excimer laser is recently used to fabricate 122 epitaxial films. As will be shown in §4, most 11 films are also grown using a KrF excimer laser. Therefore, examining the excitation energy dependence for the growth of Fe-based superconductor films by laser ablation should be important for further advances in thin film growth technology.

Hiramatsu et al. [83] reported that a distinctive phenomenon, water-induced superconductivity, occurs in undoped Sr122 epitaxial films. The black trace in Fig. 9(a) shows the $\rho$–$T$ curve of a virgin undoped Sr122 film. An anomaly in resistivity, which corresponds to a structural phase transition and antiferromagnetic ordering [84,85], can be seen at $T_{anom}$ = 204 K. On the other hand, exposing a virgin film to an ambient atmosphere at room temperature for 2 h causes $\rho$ to decrease from 25 K. As the exposure time increases, the decrease in $\rho$ becomes sharper, and finally zero resistance is observed at exposure time ≥ 4 h along with clear shrinkage of the $c$-axis length. $T_c^{onset}$ and $T_c^{zero}$ of the film exposed to air for 6 h are 25 and 21 K, respectively.

The effects of the constituents of air on the appearance of superconductivity were examined individually to determine its origin [Fig. 9(b)]. Exposure to dry nitrogen, oxygen, or carbon dioxide gas for 24 h at room temperature did not induce a superconducting transition, whereas a clear superconducting transition occurred at $T_c^{onset}$ = 25 K upon exposure to water vapor (the dew point is +13 °C). Therefore, it is revealed that water vapor or a water-related species induces superconductivity. It is noteworthy that this phenomenon occurs at a very low dew point of +13 °C (this value corresponds to 47% RH at 25 °C), indicating that undoped Sr122 is extremely sensitive to water vapor. A similar finding has been reported for undoped Ba/Sr122 single crystals [86,87], in which the strain generated during crystal growth induces the superconductivity. Subsequently, it was reported that water-related species induce superconductivity in bulk samples of 1111 [88,89] and 11 compounds [90–92] (as discussed later, there have been a few reports on 11 thin films). Thus, elucidating the interaction of water-related chemical species with the parent compounds could provide insight into the superconducting mechanism or a clue to realizing a higher $T_c$ by new doping methods such as chemical modification.

The superconducting transition shown above is invoked upon the exposure of undoped Sr122 to water vapor, and is accompanied by shrinkage of the *c*-axis length of the unit cell. Kamiya et al. [93] used a first-principles calculation to propose a model to explain *c*-axis shrinkage as well as to evaluate the structure and formation energies of water-related impurities and vacancies in Sr122. They found that the incorporation of an O, OH, or $H_2O$ molecule at an interstitial site [$I_6$ or $I_9$ shown in Fig. 10(a)] cannot explain the *c*-axis shrinkage, and a large formation energy is necessary to form an As vacancy ($V_{As}$). Their findings indicate that $V_{As}$ is not easily created in Sr122 (this may contribute to the stability and the easy thin film growth of the 122 phase). An Fe vacancy ($V_{Fe}$) is formed with a negative formation energy and can explain the *c*-axis shrinkage [Figs. 10(b) and 10(c)]. Regarding the possibility of $V_{Sr}$, the formation energy becomes negative upon the existence of $H_2O$ and the formation of $Sr(OH)_2$, and the formation of $Sr(OH)_2$ is actually detected by XRD when a Sr122 powder sample is exposed to water vapor [94]; however, the formation of $V_{Sr}$ increases the *c*-axis length. Therefore, they implied that the formation of $V_{Fe}$ is a plausible origin of *c*-axis shrinkage for Sr122 exposed to water vapor.

Note that Hanna et al. [89] recently reported that hydrogen can be incorporated as $H^-$ anions and occupy the $O^{2-}$ sites in the 1111 phase. The resulting $SmFeAsO_{1-x}H_x$ ($x = \sim 0.2$) is a superconductor with $T_c^{onset} = \sim 55$ K. Additionally, up to 40% of the $O^{2-}$ ions can be replaced by $H^-$ anions, along with electrons being supplied into the FeAs layer to maintain the charge neutrality ($O^{2-} \leftrightarrow H^- + e^-$). This result may be related to that for the superconducting Sr122 film exposed to water vapor.

### *3.2 Ba system*
Katase et al. [95] were the first to fabricate Ba122:Co epitaxial films, and they experimentally demonstrated that the Ba system is more stable than the Sr system. Their findings that Ba122:Co is more chemically stable (handling ease) and that it is much easier to grow high-quality epitaxial films than 1111 films have led to the current research activities on Ba122:Co films. Figures 11(a) and 11(b) summarize the chemical stability of Ba122:Co epitaxial films. When exposed to water vapor with dew points of +20–25 °C at 760 Torr, the Sr122:Co film rapidly begins to decompose into $Fe_2As$ [+$Sr(OH)_2$ [94]] accompanied by shrinkage of the *c*-axis in the remaining Sr122:Co phase, which alters the electrical properties such as $T_c^{zero}$ and increases normal-state resistivity. On the other hand, for Ba122:Co, exposure to water vapor does not alter the XRD pattern or the original $\rho$–$T$ curve. Also for bulk polycrystal samples, better stability of Ba122 than Sr122 is reported in refs. [22], [85], and [96], which is consistent with the thin film case. On the other hand, the origin of stability in Ba122 remains debatable. The formation energies of the impurities generated by exposure to water vapor such as $Sr(OH)_2$ and $Ba(OH)_2$ do not differ significantly. Currently, it is speculated that the size of the interstitial sites may be related to the chemical stability. The 122 structure has two interstitial sites [$I_6$ and $I_9$ shown in Fig. 10(a)], and the sizes of both sites in Ba122 are much smaller than those of Sr122, which may reduce the reactivity of Ba122 with water-related species.

Additionally, Lee et al. [97] and Iida et al. [98] reported high-quality Ba122:Co epitaxial films. Lee et al. fabricated Ba122:Co epitaxial films on four types of [001]-tilt $SrTiO_3$ (STO) bicrystal substrates, and achieved a high intragrain $J_c$ of 3 MA/cm$^2$ at 4.2 K. They first discussed the grain boundary nature of Ba122:Co, which is very important for wire/tape applications. Later in this paper, we compare their findings with the recent results reported by Katase et al. [99]. Iida et al. examined the effect of strain on $T_c$ for Ba122:Co films (Fig. 12) using four different single-crystal substrates. They concluded that STO is the best substrate because the chemical composition of each film does not fluctuate and the superconductivity properties of STO sample ($T_c^{onset}$ = 24.5 K, which is the highest $T_c$ of the Ba122:Co films reported to date with a sharp transition $\Delta T_c$ = ~1 K) are superior.

To obtain higher quality Ba122:Co epitaxial films, Lee et al. [100] proposed employing thin perovskite oxide template layers of STO or $BaTiO_3$ (BTO) with 50–100 unit cells. Figure 13 illustrates the proposed concept; they claimed that an STO/BTO buffer can effectively establish well-ordered chemical bonds between 122 films and single-crystal substrates possessing trivalent rare-earth metals such as LSAT and LAO because these buffer layers consist of the same divalent alkaline earth ions as the 122 phase. Figures 14 and 15 show the superconducting transition (the resistivity and diamagnetism) and the magnetic field/angular dependences of $J_c$ of Ba122:Co epitaxial films, respectively. In the case of bare LSAT, the normal-state resistivity is high, the transition is broad, and the shielding volume fraction is small. However, in the case of a 100 u.c. STO or 50 u.c. BTO buffer layer on an LSAT substrate, the properties drastically improved to $T_c^{onset}$ = 22.8 K and $T_c^{zero}$ = 21.5 K ($\Delta T_c$ = 1.3 K) for the STO buffer, demonstrating clear diamagnetism and a self-field $J_c > 1$ MA/cm$^2$ at 4.2 K for both buffers. $J_c$ of the film using the 50 u.c. BTO buffer was as high as 4.5 MA/cm$^2$ at 4.2 K, which is consistent with a uniform and continuous magneto-optical image. Additionally, they observed a strong $c$-axis $J_c$ peak similar to that of a low-$J_c$ Sr122:Co film [81], which is opposite to the result expected from the electronic anisotropy because the upper critical magnetic field of $B \parallel c$ is lower than that of $B \perp c$ [44,80]. The transmission electron microscopic images clearly showed strong pinning along the $c$-axis parallel to the vertically aligned nanosize defects of the secondary phases [101]. Recently, the vortex pinning, which leads to the high $J_c$, was clarified to originate from 4–5 nm BaFeO$_2$ nanopillars [102].

Iida and co-workers [103–105] proposed another effective buffer layer, metal Fe with a bcc structure, because the Fe–Fe bond length along the [110] direction for bcc Fe metal is very close to that along the [100] direction of Ba122. Therefore, even in single-crystal substrates with relatively large lattice mismatches such as LSAT (–2%) and MgO (+6%), a thin epitaxial Fe buffer layer grown as (001)[110] Fe ∥ (001)[100] LSAT/MgO contributes to coherent growth at the Fe/Ba122 interface (Fig. 16). By optimizing the thickness of the Fe buffer layer to 15 nm [104], they obtained $J_c$ of 0.45 MA/cm$^2$ at 12 K, which is an order of magnitude higher than that in their previous study [103]. In addition, the angular dependence of $J_c$ of the Ba122:Co films using the Fe buffer layer exhibited a clear $J_c$ peak when external magnetic fields were applied along $H \perp c$ (Fig. 17), which clearly differs from the trend of high-$J_c$ films with BaFeO$_2$ nanopillars [Fig. 15(b)] [100,102]. They concluded that these observations originate from intrinsic pinning due to the correlated $ab$-planes, i.e., three dimensional pinning [106].

Both the aforementioned buffer layers are electrically conductive because STO buffer layers become conductive upon the high-temperature/high-vacuum deposition of a Ba122:Co film, and Fe is a metal. Hence, it is natural to consider that these buffer layers affect the normal-state resistivity and $J_c$. Katase et al. [107] successfully fabricated high-$J_c$ Ba122:Co epitaxial films on insulating LSAT (001) single crystals without a buffer layer by employing a pure-phase laser ablation target and improving the uniformity of the substrate temperature. This method led to the first demonstration of a Josephson junction [108] and a superconducting quantum interference device (SQUID) [109] using epitaxial thin films of Fe-based superconductors (the device performance is discussed in §5). They synthesized highly purified Ba122:Co targets for laser ablation by solid-state reactions of stoichiometric mixtures of BaAs, $Fe_2As$, and $Co_2As$ via the reaction $BaAs + 0.92Fe_2As + 0.08Co_2As \rightarrow BaFe_{1.84}Co_{0.16}As_2$. In this reaction, the quality of the BaAs precursor is critical because previously employed targets contained impurities (ca. several percent) such as Fe, FeAs, $Fe_2As$, and $FeAs_2$. Therefore, they refined the preparation process of the BaAs precursor to enhance the reaction between Ba and As in an evacuated silica tube as shown in Fig. 18(a). First, large pieces of Ba metal were finely cut (1) → (2), and then crushed into small planar grains (2) → (3). This procedure was repeated until the pieces were less than ~1 mm (3) → (4). Employing a high-quality target and enhanced temperature uniformity within the substrate drastically improved the phase purity, crystallinity, and superconducting properties of the Ba122:Co epitaxial films. Figure 18(b) shows the out-of-plane XRD patterns and self-field $J_c$ of previously reported Ba122:Co epitaxial films (low-quality, LQ) and the improved films (high-quality, HQ). The HQ film shows only Ba122:Co 00$l$ and LSAT 00$l$ diffraction peaks, although the previous LQ sample exhibited an extra peak due to Fe 200 diffraction. $\Delta T_c$ of the HQ film [Fig. 18(c)] becomes sharper ($\Delta T_c$ = 1.1 K) than that of the LQ film (2.9 K). $J_c$ at 4 K [Fig. 18(d)] increases from 0.2 MA/cm$^2$ for the LQ film to 4.0 MA/cm$^2$ for the HQ film, which exhibits a sharp transition to the normal state in the $I$–$V$ curve [whereas the LQ film shows a flux-flow component due to the weak link, which is shown in the inset of Fig. 18(d)]. This $J_c$ value is comparable to that of high-$J_c$ films fabricated using STO/BTO buffer layers [100].

It should be noted that, at present, only Katase et al. have succeeded in the direct deposition of high-$J_c$ (> 1 MA/cm$^2$) Ba122:Co epitaxial films without a buffer layer. We speculate that a Nd:YAG laser may be effective for directly growing epitaxial films and introducing naturally formed pinning centers. In addition, similar high-quality epitaxial films were grown on MgO (001) single-crystal substrates with a larger in-plane lattice mismatch (+6%) than that of LSAT [110]. A detailed study on the angular dependence (Fig. 19) of $J_c$ for high-$J_c$ Ba122:Co epitaxial films was recently performed [111]. Similar to the high-$J_c$ films on STO buffer layers [100,101], these Ba122:Co films exhibit a clear $J_c$ peak around the $c$-axis direction ($\theta$ = 0), implying that correlated defects are easily introduced in Ba122:Co epitaxial films if the substrate/buffer layers differ. At higher magnetic fields of 12 and 15 T, an additional peak, which corresponds to the orientation on the $ab$-plane, can be observed. This peak is tentatively attributed to random disorder. Further studies such as microscopic texture observations and chemical analyses are necessary to clarify the origin of these strong pinning centers.

As described above, numerous researchers have studied Ba122:Co epitaxial films because they are easily fabricated compared with the 1111 system. However, $T_c$ of Ba122:Co films is only ~20 K, which is lower than that (38 K) of K-doped Ba122 [22]. Lee et al. [112] and Takeda and co-workers [113,114] successfully fabricated K-doped 122 thin films with a high $T_c$ of ~40 K. Lee et al. (Fig. 20) employed *ex situ* laser ablation, where undoped Ba122 films were initially deposited at room temperature followed by thermal annealing at 700 °C with potassium lumps in evacuated silica glass ampoules. Such a separated thermal process effectively dopes potassium, which has a high vapor pressure, to Ba122 films. They grew *c*-axis oriented films on LAO (001) and α-$Al_2O_3$ (0001) single-crystal substrates. Both the films exhibit clear superconducting transitions [e.g., $T_c^{onset}$ = ~40 K and $T_c^{zero}$ = ~37.5 K ($\Delta T_c$ = ~2.5 K)] for the film on α-$Al_2O_3$. The transition width of these films was larger than that of Ba122:Co films reported recently, but this was the first demonstration of potassium doping in Ba122 films. On the other hand, Takeda et al. (Fig. 21) used MBE to successfully fabricate two types of films (Ba and Sr systems). Reducing the As flux decreased the growth temperature of the 122 films from 540–600 °C for undoped films to 300–350 °C, and effectively doped potassium to the films. They emphasized that to realize such low-temperature growth for 122 films, potassium enhances the migration of deposition precursors during growth.

The above two sections summarized the current status of research on the Sr/Ba122 superconducting thin films and their properties. In particular, research on Ba122:Co has advanced the most significantly among the Fe-based superconductor thin films, and their values of $J_c$ have reached high levels comparable to or even exceeding those of single crystals (typical values for single crystals are on the order of 0.1 MA/cm$^2$ [115–122], although some works report a higher $J_c$ that exceeds 1 MA/cm$^2$ [123,124]). Such high-quality thin films allow various physical measurements such as terahertz spectroscopy [125–130], point-contact Andreev reflection [131,132], and superfluid density measurements [133]. Furthermore, the fabrication of K-doped 122 thin films with $T_c$ = ~40 K was reported [112–114]. If the film quality and superconducting properties such as crystallinity, $\Delta T_c$, and $J_c$ are further improved, K-doped Ba122 films should become increasingly promising candidates for thin film devices compared with 1111 films owing to the low anisotropy of K-doped Ba122 compound [44].

*3.3 Grain boundaries*

For high-$T_c$ cuprates, a critical issue in the development of superconducting wires/tapes for electric power transmission is the 'grain boundary (GB)' issue. The grains of high-$T_c$ cuprates must be highly textured to prevent the deterioration of $J_c$ across misaligned GBs because $J_c$ strongly depends on the misorientation angle of GBs ($\theta_{GB}$). For example, a fundamental study on the intergrain $J_c$ ($J_c^{BGB}$) for a representative cuprate, YBa$_2$Cu$_3$O$_{7-\delta}$ (YBCO), was conducted using several types of bicrystal substrates [134]. Significantly misaligned adjacent grains cause $J_c^{BGB}$ to decay exponentially as a function of $\theta_{GB}$ from 3 to 40° [135]. Therefore, to produce YBCO superconducting wires/tapes with a high $J_c$, well-aligned buffer layers with a small in-plane misalignment of $\Delta\phi < 5°$ on polycrystalline (nonoriented) metal substrates must be inserted using an ion-beam-assisted deposition (IBAD) technique [136] or rolling-assisted biaxially textured substrates (RABiTS) [137]. Although recent progress in buffer-layer technology has led to the achievement of $\Delta\phi < 5°$, which provides a self-field $J_c$ of several MA/cm$^2$ at 77 K [138], fabricating such buffer layers is time-consuming and expensive. New high-$T_c$ superconducting materials with a more gradual $J_c^{BGB}(\theta_{GB})$ dependence allow a simple, low-cost process and provide more flexibility in superconductor power lines.

Lee et al. [97] were the first to examine the transport properties of iron pnictide superconductors through bicrystal GBs (BGBs) using Ba122:Co epitaxial films grown on [001]-tilt STO bicrystal substrates with four different $\theta_{GB}$ of 3, 6, 9, and 24°. Figure 22 shows the $J_c^{BGB}(\theta_{GB})$ curve at 12 K under a magnetic field ($B$) of 0.5 T. $J_c^{BGB}$ decreases by an order of magnitude as $\theta_{GB}$ increases from 3 to 24°, and the critical angle ($\theta_c$), which is defined as the $\theta$ where the transition from a strong link to a weak link occurs, is estimated to be 3–6°. In addition, $J_c^{BGB}$ rapidly decreases from the intragrain $J_c$ ($J_c^{Grain}$) value in the low-$B$ region, but these values become comparable at higher fields (Fig. 23). The weak-link behaviors observed for Ba122:Co BGBs are similar to those observed for YBCO BGBs.

On the other hand, Katase et al. [99] recently reported that $J_c^{BGB}$ of Ba122:Co has a gentler $\theta_{GB}$ dependence than that of YBCO, as examined using high-$J_c$ Ba122:Co epitaxial films on [001]-tilt LSAT and MgO bicrystal substrates with $\theta_{GB}$ = 3–45°. Figure 24 shows the $J_c^{BGB}(\theta_{GB})$ curves in a self-field at 4 and 12 K. Ba122:Co has a high $J_c^{Grain}$ of ~2 MA/cm$^2$ at 4 K, and the deterioration of $J_c^{BGB}$ due to the tilted GBs is negligible at $\theta_{GB}$ lower than $\theta_c = 9°$, which is twice as large as $\theta_c = 5°$ for YBCO BGBs. When $\theta_{GB} > \theta_c$, $J_c^{BGB}(\theta_{GB})$ exhibits nearly exponential decay given by $J_c^{BGB}(\theta_{GB}) = J_{c0}\exp(-\theta_{GB}/\theta_0)$, but the slope coefficient of $\theta_0 = 9°$ is larger than $\theta_0 = 4°$ for YBCO. These observations indicate that the decay of Ba122:Co is gentler than that of YBCO in the weak-link regime. $J_c^{BGB}$ decreases to ca. 5% as $\theta_{GB}$ increases from 9 to 45°, but the $J_c^{BGB}$ of the Ba122:Co BGBs exceeds that of the YBCO BGBs at $\theta_{GB} \geq 20°$ at 4 K because of the gentler decay for Ba122:Co. $J_c^{BGB}(B)$ with low values of $\theta_{GB} = 3$ and 4° exhibits similar behavior to $J_c^{Grain}(B)$, but $J_c^{BGB}(B)$ with a larger $\theta_{GB}$ shows a more rapid decrease even in the low $B$ region (Fig. 25).

Note that Lee et al. reported a smaller $\theta_c$ of 3–6° [97]. The difference between the studies by Lee et al. [97] and Katase et al. [99] may be film quality because Katase et al. reported that the self-field $J_c^{Grain}$ at 12 K is one order of magnitude higher than that reported by Lee et al. In addition, this difference (i.e., $\theta_c = 3$–6 or 9°) is large in terms of practical applications because it is much easier to form a buffer layer with $\Delta\phi \leq 9°$ than the buffer layer required for cuprates, as evidenced by the fact that the rather small $\Delta\phi$ of < 5° has been achieved by only two groups [139,140] using additional MgO or $CeO_2$ buffer layers. Therefore, the large $\theta_c$ allows a simpler and lower-cost production process to be used to produce superconducting wires and tapes.

### 4. 11 Thin Films

Thin films of iron chalcogenides were reported prior to the first report on superconductivity. Although these early reports focused on the electrical and magnetic properties, superconductivity was not observed [141–143].

Wu and co-workers [144] were the first to report 11 superconducting films in a review article as an unpublished work after they observed superconductivity in a bulk 11 compound [24]. Their first paper was published in ref. [145]. They reported that the preferential orientation and $T_c^{onset}$ of 11 films grown by laser ablation using a KrF excimer on MgO (001) single crystals strongly depend on the growth temperature and film thickness (Fig. 26). $FeSe_{1-x}$ films have (001) and (101) preferential orientations when grown at 320 °C (LT) and 500 °C (HT), respectively. In particular, $T_c^{onset}$ depends strongly on thickness for the LT films, but no obvious dependence was observed in the cases of HT growth. The HT film has a maximum $T_c^{onset}$ of 10 K and a $T_c^{zero}$ of 4 K ($\Delta T_c$ = 6 K), but a 600-nm-thick FeSe$_{0.5}$Te$_{0.5}$ film has a higher $T_c^{onset}$ of 14 K and a $T_c^{zero}$ of 12 K ($\Delta T_c$ = 2 K) [144]. Furthermore, when XRD measurements were performed at low temperatures, they observed that the structural distortion of the LT films was strongly suppressed in thinner films but recovered gradually as film thickness increased. Such a distortion was not observed in HT films. Therefore, they concluded the low-temperature structural distortion is correlated with the occurrence of superconductivity in 11 thin films.

At almost the same time, Han et al. [146] reported superconducting FeSe$_x$ ($x$ = 0.80–0.92) films grown on STO, LSAT, and LAO (001) substrates by laser ablation with a XeCl excimer laser ($\lambda$ = 308 nm). The resulting films exhibited a strong out-of-plane $c$-axis crystallographic orientation. For a ca. 200-nm-thick $x$ = 0.88 film on an LAO single-crystal substrate, they obtained a maximum $T_c^{onset}$ of 11.8 K and a $T_c^{zero}$ = 3.4 K ($\Delta T_c$ = 8.4 K) (Fig. 27). Soon after, Si et al. [147] and Bellingeri et al. [148] reported Te-substituted 11 (composition: FeSe$_{0.5}$Te$_{0.5}$) superconducting epitaxial films with higher $T_c$ values (~17 K) and a shrunken $c$-axis. Since these reports on 11 superconducting films in early 2009, many groups have fabricated superconducting 11 films [114,149–168]. However, particularly, in the initial stage of this research, there were few reports of films with zero resistivity, demonstrating the difficulty in producing superconducting 11 thin films owing to the complicated phase diagram [169].

However, Han et al. [155] performed a noteworthy study; an FeTe film, which is an end member of the 11 phase and whose bulk sample does not exhibit superconductivity without replacement [170–172], shows superconductivity at $T_c^{onset}$ = 13.0 K, $T_c^{zero}$ = 9.1 K ($\Delta T_c$ = 3.9 K) owing to tensile strain (Fig. 28). A critical thickness of ~90 nm is necessary to realize the maximum $T_c$. Clear diamagnetism is observed along with a shielding volume fraction of 22%. Therefore, the surface and/or impurity effect may be negligible, but they commented that the possible inhomogeneity in the films may cause incomplete diamagnetism (not 100%) and the large $\Delta T_c$. In addition, the obtained self-field $J_c$ remained at 0.067 MA/cm$^2$ at 2 K, suggesting weak-link behavior due to a granular structure.

In 2010, Bellingeri et al. [158] reported the highest $T_c$ among the 11 thin films to date: $T_c^{onset}$ = 21 K and $T_c^{zero}$ = 19 K ($\Delta T_c$ = 2 K). This $T_c$ value is higher than that of the bulk 11 phase at ambient pressure (~14 K) [172–174] and is comparable to that of Sr/Ba122:Co thin films (discussed in §3). They employed a target with a chemical composition of FeSe$_{0.5}$Te$_{0.5}$ and a laser ablation technique using a KrF excimer laser. They grew 11 films on three types of substrates [LAO, STO, and yttria-stabilized zirconia (YSZ)] with various thicknesses and found that significant and systematic changes do not occur in the $c$-axis lengths. In contrast, the $a$-axis lengths [Fig. 29(a)] decrease as the thickness increases and reach a minimum at 200 nm, leading to the highest $T_c$ of 21 K [Figs. 29(b) and 29(c)]. Additionally, Imai et al. [160] examined eight types of single-crystal substrates for $FeSe_{0.5}Te_{0.5}$ epitaxial growth by laser ablation and reported that the in-plane orientation and $c/a$ ratio, but not the lattice mismatch, are critical for inducing the superconductivity in 11 films.

As shown above, particularly for 11 films, many studies on stress versus $T_c$ have performed. Huang et al. [161], who studied the local structure in epitaxial 11 films, indicated that the geometry of an $Fe(Se,Te)_4$ tetrahedron, which is represented by the chalcogen height ($h_{Se/Te}$) in the iron chalcogenide layer, is critical for inducing superconductivity. Their finding is consistent with a theory proposed by Kuroki et al. in 2009 [175]. $FeSe_{0.5}Te_{0.5}$ thin films with thicknesses of ~400 nm have been grown on MgO (001) by laser ablation using a KrF excimer laser. The growth temperature, which plays a key role in epitaxy and superconductivity, was varied from 180 to 500 °C, and different types of stress were induced in the films. As the $c$-axis length increases, $z$ (the fractional coordinate of the chalcogen in the 11 phase) and $h_{Se/Te}$ monotonically decrease, but the Se−Se length increases, indicating that the tetrahedron undergoes compressive distortion and that the FeSe/Te plane becomes thinner. As shown in Fig. 30, the superconducting transition initially appears at $\Delta h_{Se/Te}$ ($= c\Delta z + z\Delta c$) $\approx$ 0.05 Å. Upon further decreasing $\Delta h_{Se/Te}$, $T_c$ increases and is maximized at $\Delta h_{Se/Te} \approx 0$ Å. Therefore, they concluded that $T_c$ for the $FeSe_{0.5}Te_{0.5}$ thin films is strongly linked to $h_{Se/Te}$ but not to epitaxy.

Oxygen-incorporating effects in FeTe thin films similar to those in 11 bulk samples [91,92] were reported by Si et al. [156] and Nie and co-workers [157,166]. Si et al. grew FeTe epitaxial films on STO substrates by laser ablation. When grown in a vacuum of $2\times10^{-6}$ Torr, a gentle drop in resistivity begins at ~10 K. However, films grown in an oxygen atmosphere (~$1\times10^{-4}$ Torr) clearly exhibit zero resistance (Fig. 31). Nie et al. examined the air-exposure effect of FeTe films grown on MgO and found that annealing in oxygen at 100 °C induces superconductivity in a FeTe film (Fig. 32). These reports indicate that oxygen is associated with the emergence of superconductivity in FeTe. Similarly to undoped Sr122 films [83], such an oxygen/water-induced effect may lead to a novel effective chemical modification technique of Fe-based superconductors once the detailed mechanism is clarified.

This section summarized the reports on 11 thin films. Among the Fe-based superconductors, the 11 phase has the simplest chemical composition, and the incorporation of impurities with high vapor pressures such as F and K is unnecessary to induce superconductivity. These features allow various thin film fabrication processes to be applied. Not only laser ablation but also other methods such as MBE [114] and metal-organic chemical vapor deposition (MOCVD) [165] have been used for thin film growth for this system. Although most of the $T_c$ values are lower than those of other Fe-based superconductor films, 11 thin films have been reported to possess a distinct strain effect on $T_c$. This result may correspond to the drastic increase in $T_c$ of the 11 phase to up to 37 K under high pressure [176–179]. The $J_c$ values of 11 single crystals are lower than those of other Fe-based superconductors; typical values are ~0.1 MA/cm$^2$ at low temperatures [180–183] (The highest value reported is 0.6 MA/cm$^2$ at 1.8 K [182].). For 11 thin films, reports on their $J_c$ values are limited to a few papers [159,167,168]. Most of the values of $J_c$ are lower than those of 11 single crystals and 1111/122 thin films. However, Eisterer et al. [167] recently reported FeSe$_{0.5}$Te$_{0.5}$ thin films exhibiting $J_c$ on the order of 0.1 MA/cm$^2$. Further improvements are necessary to apply 11 thin films to superconducting devices.

## 5. Superconducting Thin Film Devices
### 5.1 Thin film Josephson junctions

Josephson junctions using Fe-based superconductors are useful for fundamental experiments, including symmetry determination of the superconducting order parameter and clarification of potential applications to superconducting electronics. Josephson effects using single crystals of 122 superconductors were observed early on in Fe-based superconductor research. In 2008, a Josephson junction of Pb/Ba122:K was fabricated, and phase-sensitive measurements on a corner junction of Ba122:Co were performed using a Pb counter electrode [184,185]. The results indicate that these superconductors have $s$-wave symmetry. Furthermore, the Josephson effect was observed in junctions formed between electron-doped Sr122:Co and hole-doped Ba122:K [186]. In the case of the 1111 phase, the half-flux quantum effect, which is due to a sign change in the order parameter, was observed in a superconducting loop composed of a fluorine-doped Nd1111 polycrystalline bulk and Nb wires [187]. Additionally, $c$-axis transport measurements on a Pr1111 single crystal [75] clarified that intrinsic Josephson junctions with insulating barriers are formed in a Pr1111 crystal.

Josephson junctions based on epitaxial thin films are necessary to realize superconducting electronics applying Fe-based superconductors. However, as shown in §2, fabricating high-quality 1111 thin films still remains difficult. Therefore, it is difficult to fabricate 1111-thin-film-based superconductor devices [188]. Under such circumstances, Katase et al. [108] were the first to successfully fabricate a thin-film-based Josephson junction using high-quality Ba122:Co epitaxial films on [001]-tilt LSAT bicrystal substrates with $\theta_{GB} = 30^o$ (Fig. 33). 10-μm-wide Ba122:Co microbridges across BGBs (BGB junctions) were formed to observe the Josephson effect, and microbridges without a BGB (non-BGB junctions) were used to compare the $J_c$ values with and without a BGB junction. For the BGB junctions, the shape of the I–V curve at B = 0 mT displays resistively-shunted-junction (RSJ) type behavior without hysteresis. The estimated normal-state resistance ($R_N$) and $R_NA$ product (A is the cross-sectional area of the junction) of the BGB junctions at 10 K are 0.012 Ω and $3.0 \times 10^{-10}$ Ωcm$^2$, respectively. On the other hand, $I_c$ is clearly suppressed by a weak magnetic field [B = 0.9 mT, Fig. 34(a)]. The large $I_c$ modulation of 95% indicates that the Josephson current is responsible for most of the supercurrent through the BGB junction. Examination of the B dependence [Fig. 34(b)] demonstrated that $I_c$ decreases rapidly and monotonically, probably due to the large junction width. In addition, the position of the maximum $I_c$ shifts from the zero magnetic field, indicating that an ideal Fraunhofer pattern is not observed. Instead, short-period oscillations due to the trapped flux emitted from external electronic devices occur. Throughout the temperature range, $J_c$ of the BGB junctions [Fig. 35(a)] is about 20 times smaller than that of the non-BGB junctions, implying that BGBs work as weak-link GBs. The $I_cR_N$ product increases almost linearly with decreasing temperature, and the $I_cR_N$ product at 4 K is 55.8 μV [Fig. 35(b)]. The two orders of magnitude smaller $I_cR_N$ product compared with that of a YBCO BGB junction originates from the metallic nature of Ba122:Co BGB junctions.

Later, Schmidt et al. [189] reported a multi layered (superconductor-normal metal-superconductor, SNS) Josephson junction using a Ba122:Co film and PbIn counter electrodes with a 5-nm-thick Au barrier layer (Fig. 36). The junction area is $30 \times 30$ μm$^2$. As shown in Fig. 37, the R–T curve exhibits two steps at temperatures of 7.2 and 15 K, which correspond to the $T_c$ values of PbIn and Ba122:Co, respectively. Depending on the current bias direction, the I–V characteristic of the hybrid junction is slightly asymmetric, but the I–V curve has an RSJ-like nonlinear shape with hysteresis. The $I_cR_N$ product and $J_c$ are 18.4 μV and 39 A/cm$^2$ at 4.2 K, respectively. The I–V curves under microwave irradiation at frequencies of 10–18 GHz clearly display multiple Shapiro steps, confirming the Josephson effect (Fig. 38).

To date, only two groups have reported Josephson junctions using Fe-based superconductor thin films. One was formed at a bicrystal grain boundary, and the other employed an SNS structure. Both of these studies used Ba122:Co films, which have the highest $J_c$ values among Fe-based superconductor films.

*5.2 SQUID*

One of the most important applications of Josephson junctions is the SQUID, which has been recognized as a highly sensitive magnetic sensor. Additionally, the SQUID is a useful tool for evaluating thin film qualities and junction barriers and for determining superconducting symmetry. Katase et al. [109] successfully fabricated a dc-SQUID composed of a superconducting loop with two BGB junctions in a Ba122:Co epitaxial film on a bicrystal substrate. A SQUID loop with a slit area of $18 \times 8$ μm$^2$ and a narrow ring width of 3 μm was formed across the BGB and involved two Josephson junctions (Fig. 39). Periodic voltage modulation of $\Delta V = 1.4$ μV was observed in the voltage–flux ($V$–$\Phi$) characteristics of the Ba122:Co dc-SQUID measured at 14 K (inset in Fig. 40).

Furthermore, a flux-locked loop (FLL) circuit was employed to evaluate the flux noise $S_\Phi^{1/2}$ spectrum of the dc-SQUID (Fig. 40). The $S_\Phi^{1/2}$–$f$ spectrum exhibits $1/f$ noise in the low-$f$ region (< 20 Hz) and white noise with a constant $S_\Phi^{1/2}$ of $1.2 \times 10^{-4}$ $\Phi_0$/Hz$^{1/2}$ in the $f$ > 20 Hz region. The high $S_\Phi^{1/2}$ level includes an equivalent input noise in an FLL circuit, which obscures the intrinsic $S_\Phi^{1/2}$ ($S_{\Phi,\text{intrinsic}}^{1/2}$) of the dc-SQUID owing to the low $\Delta V$. However, if the noise is excluded by a preamplifier, $S_{\Phi,\text{intrinsic}}^{1/2}$ is estimated to be $7.8 \times 10^{-5}$ and $4.2 \times 10^{-4}$ $\Phi_0$/Hz$^{1/2}$ in the white noise region and at $f = 1$ Hz, respectively. The $S_\Phi^{1/2}$ level of the Ba122:Co dc-SQUID is at least ten times higher than $S_\Phi^{1/2} = 1.0 \times 10^{-5}$ $\Phi_0$/Hz$^{1/2}$ of a YBCO dc-SQUID with $\Delta V = 10$–$20$ μV at 77 K [190,191]. The operation temperature of this Ba122:Co dc-SQUID must be near $T_c$ because $I_c$, which must be smaller than 200 μA owing to the requirement of dc-SQUID operation using the FLL circuit, rapidly increases with decreasing temperature. In addition, $V_\Phi$ is low owing to the low $R_N$, which is attributed to the metallic nature of normal-state Ba122:Co. Consequently, the high measurement temperature and low $V_\Phi$ are responsible for the high noise level. Therefore, artificial barriers with large junction resistances, such as superconductor-insulator-superconductor (SIS) junction structures, are necessary to realize practical SQUIDs with iron-pnictide superconductors.

However, this demonstration of a dc-SQUID using BGBs should enable an innovative probe technique to examine pairing symmetries.

### 5.3 Wires and tapes

The wire fabrication of Fe-based superconductors based on 1111, 122, and 11 phases by the powder-in-tube (PIT) method has been examined since 2008. Gao et al. first demonstrated superconducting wires of F-doped La1111 with a Ti-buffered Fe sheath [192] and F-doped Sm1111 with a Ta sheath [193] via an *in situ* PIT process with high values of $T_c^{onset}$ of 24.6 and 52 K, respectively. Qi et al. [194] also realized Nb-sheathed Sr122:K wires with $T_c^{onset}$ of 35.3 K. However, neither group reported the transport $J_c$, presumably because of the severe reaction between the sheath and superconducting cores and/or the low sintering density of the resulting wires. Mizuguchi et al. [195] first realized a supercurrent-carrying wire using an Fe-sheathed $FeSe_{1-x}Te_x$ wire via a unique *in situ* PIT process; they reacted the Fe sheath with a Se/Te precursor to form an $FeSe_{1-x}Te_x$ superconducting core. On the other hand, Wang et al. found that a Ag buffer layer inhibits the reaction between the Fe sheath and core materials, and they observed supercurrents in Sr122:K [196] and Sm1111:F wires [197] fabricated by the *in situ* PIT process. Furthermore, they clarified that the addition of Ag to Sr122:K superconducting cores increases $J_c$ to 1.2 kA/cm$^2$ at 4.2 K. Qi et al. [198] also fabricated Sr122:K superconducting wires with a Ag/Fe sheath by an *ex situ* PIT process and realized a higher $J_c$ of 3.8 kA/cm$^2$ in Pb-added Sr122:K superconducting wires. These results indicate that Ag or Pb addition can effectively realize high-$J_c$ wires. Indeed, among recent reports [192–204], Togano et al. [203] revealed wires with the highest self-field $J_c$ (10 kA/cm$^2$ at 4.2 K), which was fabricated using Ba122:K and Ag addition by *ex situ* PIT. However, $J_c$ in superconducting wires does not satisfy the properties required for practical applications. Therefore, issues such as the reduction of the segregation of the non-superconducting phase at GBs, higher sintering density of superconducting cores, and higher texturing of crystallites are still challenging and must be overcome.

Coated conductors grown on textured metal-tape substrates may be useful for overcoming such problems because high $J_c$ values of above 1 MA/cm$^2$ have been reported for Ba122:Co thin films on single-crystal substrates. Properties close to those of thin films on single crystals may be realized on metal tape substrates if similar crystal quality, i.e., purity, orientation, etc., is achieved. However, some issues must be resolved. (Q1) Are well-known metal tape substrates applicable? (Q2) What type of tape substrate is the most suitable? Iida et al. [205] answered Q1; they fabricated

Ba122:Co films on thin Fe buffer layer / IBAD-MgO tape substrates and demonstrated a high self-field $J_c$ of over 0.1 MA /cm$^2$ at 8 K (Fig. 41). Their findings indicate that IBAD-MgO tape substrates are applicable to Fe-based superconductor thin films. However, the superconducting properties are slightly inferior to those of thin films on single-crystal substrates and the ferromagnetic Fe buffer layer may cause problems in practical applications.

On the other hand, Q2 has not yet been answered; i.e., it is not clear whether MgO is the most suitable substrate material for coated conductor applications using Fe-based superconductors. Therefore, Katase et al. [206] tried to fabricate Ba122:Co thin films on other tape substrates used for cuprates such as a top CeO$_2$ or LaMnO$_3$ layer on IBAD MgO/Gd$_2$Zr$_2$O$_7$/Hastelloy tapes [140]. However, only MgO could be used for Ba122:Co growth because the CeO$_2$ layer was reduced to Ce$^{3+}$ during the high-temperature growth of Ba122:Co under a vacuum of ~1×10$^{-5}$ Pa, and consequently, nonoriented Ba122:Co films were produced. They obtained similar results in the case of the LaMnO$_3$ top layer. Therefore, it is tentatively concluded that MgO is an appropriate substrate for Ba122:Co tape applications.

Additionally, Katase et al. [206] attempted to fabricate Ba122:Co films on IBAD-MgO substrates with the expectation of realizing a high-$J_c$ Ba122 tape because a high-$J_c$ Ba122:Co film can be directly grown on MgO without a buffer layer if a single-crystal substrate is employed. The IBAD-MgO substrates consisted of a homoepitaxial MgO layer/IBAD-MgO layer/Y$_2$O$_3$ buffer layer/Hastelloy C276 polycrystalline tape [139]. Figure 42(a) shows the $\rho-T$ curves of Ba122:Co films on IBAD-MgO and single-crystal MgO without a buffer layer using a Nd:YAG laser ablation system. Although three types of IBAD-MgO layers with $\Delta\phi_{MgO}$ between 5.5 and 7.3° were employed, the in-plane orientation of the Ba122:Co layer on IBAD-MgO was $\Delta\phi_{Ba122\,103}$ = ~3°. The superconducting properties are similar for each type of layer, but a broadened resistivity drop occurred from 26 K in the case of the tape substrate, probably because the films have an inhomogeneous Co content due to the relatively inhomogeneous substrate temperatures. The $I-V$ characteristics at 2 K for Ba122:Co on IBAD-MgO with $\Delta\phi_{MgO}$ = 7.3, 6.1, and 5.5° [inset of Fig. 42(a)] show a sharp transition, indicating that granularity is not an issue (i.e., no weak-link behavior). Values of self-field $J_c$ of 3.6, 1.6, and 1.2 MA/cm$^2$, which are higher than that reported by Iida et al. [205], were obtained for $\Delta\phi_{MgO}$ = 7.3, 6.1, and 5.5°, respectively ($J_c$ remained above 1 MA/cm$^2$ at 10 K). Because Ba1222:Co exhibits $\Delta\phi_{Ba122\,103}$ = ~3° on IBAD-MgO, the $J_c$ value, which originates from the high critical angle $\theta_c$ = ~9° [99], was almost the same as that on MgO single crystal. Both $J_c(B)$ curves almost monotonically decreased as $B$ increased [Fig. 42(b)]. However, the decay of $J_c(B)$ of the Ba122:Co tape was gentler than that on MgO single crystal, implying that the Ba122:Co tape naturally introduces strong *c*-axis pinning centers. We anticipate that further improvements may lead to the realization of a high-$J_c$ coated conductor using Ba122:Co.

### 6. Brief Summary of the Latest Results

During the submission and reviewing process of this paper, thin film research on Fe-based superconductors has continued to achieve new progress. Here, we briefly summarize the latest results.

Holzapfel's group [207] fabricated a Sm1111 film with $T_c^{onset}$ = 34 K and discussed the superconducting gaps via point-contact Andreev-reflection measurements. Ikuta's group [208,209] demonstrated that Ga, which is essential for growing Nd1111 epitaxially, effectively gathers and removes excess fluorine during MBE growth. Some effective fluorine sources for superconducting 1111 thin films such as NdOF [208], $FeF_3$ ($FeF_3 \rightarrow FeF_{3-x} + xF$) [210], and $SmF_3$ [211] were recently proposed by Ikuta and Naito's groups.

Regarding 122 compounds, a high-$J_c$ Ba122:Co film exhibiting above 1 MA/cm$^2$ was also demonstrated by Rall et al. [212], and Döring et al. [213] fabricated an SNS junction using a Ba122:Co epitaxial film as a base electrode. Iida et al. [214] proposed a new buffer layer, $MgAlO_4$, which is effective between an Fe layer and an LSAT substrate. In addition to Co and K doping, other doping modes of Ba122 films such as indirect P-doping at the As site [215] and rare-earth doping at the Ba site [216] have been reported. These indirect doping methods have led to higher $T_c$ values than those obtained by direct cobalt-doping methods in bulk samples, mainly because indirect doping modes prevent the incorporation of disorder. In particular, rare-earth doping at the Ba site appears to be favorable for thin film growth because rare-earth elements have low vapor pressures. Saha et al. [217] recently demonstrated a high $T_c$ of 47 K in the Ca122 system. The maximum $T_c$ value of bulk rare-earth-doped Ca122 [217–220] is 49 K (though the origin of such high-$T_c$ is still under debate) [219]. However, the maximum $T_c$ value of rare-earth-doped Ba122 films is almost the same as that of Ba122:Co films, implying that charge polarity is important to induce superconductivity in Ba122 and that indirect or direct doping does not dominate the superconductivity [216].

11 films have also been studied actively [221–227]. Higher quality films are now being produced using laser ablation [222,226]. As discussed in §4, the strain effect observed in 11 films is an interesting topic. However, Hanawa et al. [221] examined the effect of the substrate on the transport; their findings imply that lattice mismatch does not affect the structural and superconducting properties, and that oxygen penetration from the substrate results in poor superconductivity. Additionally, Song et al. [224] reported high-quality 11 films grown by MBE. Employing these high-quality films on SiC (0001), they investigated the electron-pairing mechanism using scanning tunneling microscopy and spectroscopy [225]. Moreover, 11 films were grown for the first time using a sputtering technique [223]. Although films grown by sputtering seem to be inferior to those grown by laser ablation, the ability to grow films by sputtering should lead to future applications of 11 films owing to financial reasons. Gooch et al. [227] examined the effect of pressures of up to 1.7 GPa on 11 films and reported that, similarly to bulk samples, $T_c$ increases as the pressure increases, but unfortunately, it is difficult to discuss their results on the basis of local structural parameters under pressure (data collection for thin films under high pressure is difficult).

For future practical applications, 1111 [228,229] and 11 [230,231] wires have been reported. Although the $J_c$ values in these reports remain of 0.1–1 kA/cm$^2$ order, Ba122:K wires [203] have the highest $J_c$ among the Fe-based superconducting wires. On the other hand, Si et al. [232] reported high-$J_c$ 11 films on IBAD-MgO tapes. Their $J_c$ values on the order of 10$^5$ A/cm$^2$ are comparable to that of Ba122:Co reported by Iida et al. [205] but lower than that (1.2–3.6 MA/cm$^2$) of Ba122:Co reported by Katase et al. [206]. It is noteworthy that $J_c$ of their 11 film still remains above 10$^4$ A/cm$^2$ in a high magnetic field of 25 T, reflecting the high upper critical magnetic field.

## 7. Summary and Future Prospects

Herein we overviewed thin films and related device fabrications for Fe-based superconductors. More than three years have elapsed since the discovery of superconductivity at 26 K in F-doped LaFeAsO. Table I summarizes particularly important properties of the thin films reviewed in this paper. The current status of this research area is summarized as follows:

(1) Although five types of parent compounds, which contain a square iron lattice as a common building block, have been reported to date, thin film fabrication has been restricted to only three materials (1111, 122, and 11 phases).

(2) The Nd1111 system exhibits the highest $T_c$ of 56 K. In the 122 system, the maximum $T_c$ for K-doped 122 films is ~40 K. The values of $T_c$ of these two systems are almost the same as those of the bulks. It is noteworthy that the maximum $T_c$ for a 11 thin film is 21 K, which is higher than that (14 K) of the bulk.

(3) Ba122:Co films with $T_c$ = ~20 K have been actively investigated, and a high $J_c$ over 1 MA/cm$^2$ has been realized.

(4) Superconducting devices such as thin-film Josephson junctions (bicrystal and SNS junctions) and a SQUID utilizing bicrystal Josephson junctions have been demonstrated using Ba122:Co films.

(5) The critical angle ($\theta_c$) of the transition from a strong link to a weak link for Ba122:Co is as high as 9°, which is twice that of cuprates ($\theta_c$ = ~5°).

(6) Only 11 films display a clear strain effect due to lattice mismatch between the substrate and material.

On the basis of the above summary, we see four future prospects.

(1) The strain effect should be present in other systems besides the 11 system. In particular, the strain effect in 1111 films may raise $T_c$ to over 56 K.

(2) Ba122:Co is expected to be a high-$J_c$ coated conductor owing to its high $\theta_c$, large upper critical magnetic field, and low anisotropy.

(3) The elucidation of effective pinning centers must be realized to achieve a higher $J_c$ because the present pinning centers appear to be introduced naturally during the thin film growth process.

(4) Exploration of the distinct properties of thin films is desirable from both experimental and theoretical viewpoints. Recently a prediction that a spin-nematic ordering structure is stabilized in an ultrathin epitaxial layer has been proposed [233].

We expect that challenges toward the realization of the above four subjects will lead to exciting innovations in Fe-based superconductors and spark increased research efforts. Finally, we would like to recommend other recent reviews on the applications of Fe-based superconductors [234–239], which are written from different viewpoints.

**Acknowledgments**

This work was supported by the Japan Society for the Promotion of Science (JSPS), Japan, through the "Funding Program for World-Leading Innovative R&D on Science and Technology (FIRST Program)". One of the authors (H. Hiramatsu) acknowledges financial support by the 16th research grant from the Seki Foundation for the Promotion of Science and Technology.

Table I. Summary of superconducting properties (chemical stability, $T_c^{onset}$, and $J_c$) for practical applications of Fe-based superconducting thin films along with the growth method. LA and TA in the growth method denote laser ablation and thermal annealing, respectively.

| Phase | Composition: dopant | Growth method | Chemical stability | $T_c^{onset}$ (K) | $J_c$ (MA/cm$^2$) | Ref. |
|---|---|---|---|---|---|---|
| 1111 | La1111:F | LA+TA | Highly stable | 28 | 0.1 | 67, 69 |
|  | (Nd/Sm)1111:F | MBE | Highly stable | 48–56 | – | 71–73 |
| 122 | Ba122:Co | LA | Stable | 20–24 | 0.5–4 | 97, 100, 104, 107 |
|  | (Ba/Sr)122:K | LA+TA or MBE | Unstable | 38–40 | – | 112, 113 |
| 11 | FeSe$_{0.5}$Te$_{0.5}$ | LA | Stable | 21 | – | 158 |
|  | FeSe$_{0.5}$Te$_{0.5}$ | LA | Stable | 19 | ~0.8 | 167 |

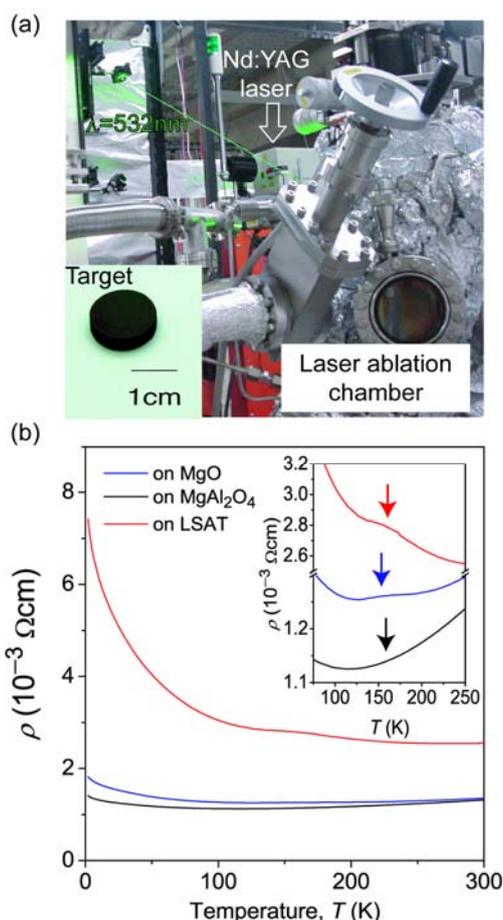

Fig. 1. (Color online) (a) Photographs of the laser ablation system using a Nd:YAG laser and a F-doped LaFeAsO laser ablation target. (b) $\rho$–$T$ curves of La1111 epitaxial films on three types of (001)-oriented single-crystal substrates. The inset shows magnifications of the curves around 150 K. Reprinted from H. Hiramatsu et al.: Appl. Phys. Lett. **93** (2008) 162504 [61]. Copyright 2008 American Institute of Physics.

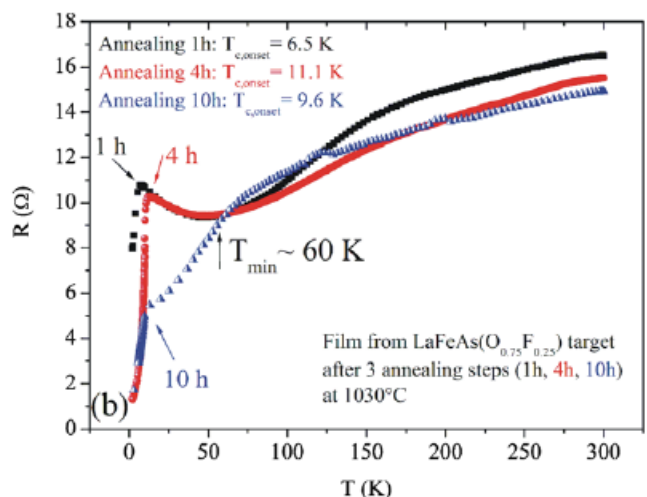

Fig. 2. (Color online) Temperature dependence of resistance of La1111 films on LAO (001) single-crystal substrates grown by a method combining laser ablation for deposition with thermal annealing for crystallization. Reprinted from E. Backen et al.: Supercond. Sci. Technol. **21** (2008) 122001 [64]. Copyright 2008 Institute of Physics Publishing.

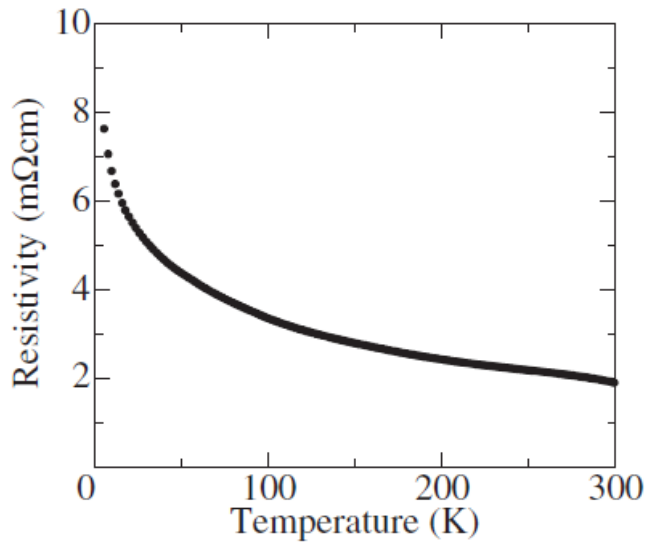

Fig. 3. Temperature dependence of resistivity of Nd1111 epitaxial film grown on GaAs (001) substrate by MBE. Reprinted from T. Kawaguchi et al.: Appl. Phys. Express **2** (2009) 093002 [65]. Copyright 2009 Japan Society of Applied Physics.

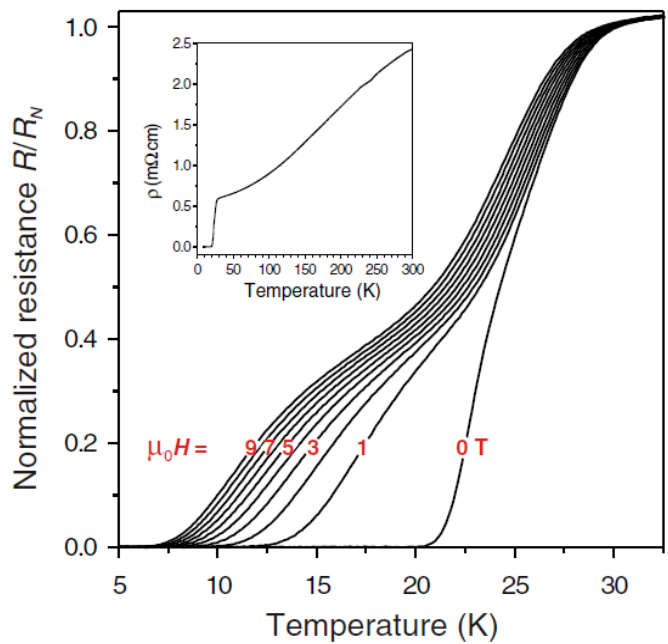

Fig. 4. (Color online) Temperature and external magnetic field dependences of normalized resistance of La1111 film on LAO (001) substrate grown by a method in which laser ablation for film deposition and thermal annealing for crystallization are combined. The inset shows a wide-range view between 2 and 300 K. Reprinted from S. Haindl et al.: Phys. Rev. Lett. **104** (2010) 077001 [66]. Copyright 2010 American Physical Society.

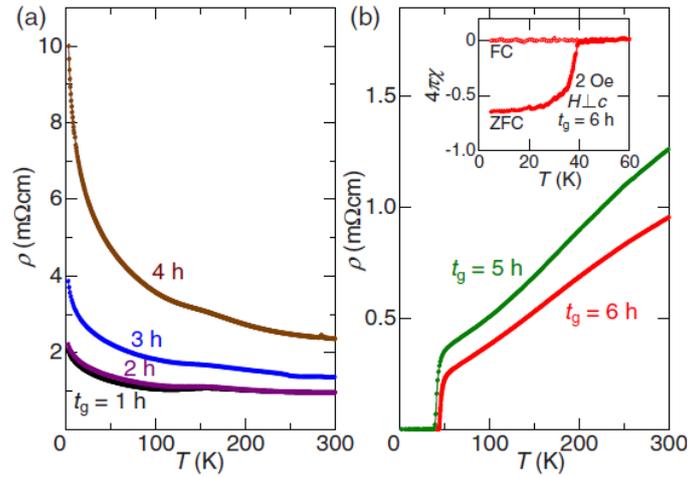

Fig. 5. (Color online) Change in $\rho$–$T$ curves of Nd1111 epitaxial films grown on GaAs (001) by MBE as a function of growth time ($t_g$) (a: $t_g \leq 4$ h, b: $t_g \geq 5$ h). The inset of (b) shows the temperature dependence of the magnetic susceptibility of the $t_g = 6$ h film. Reprinted from T. Kawaguchi et al.: Appl. Phys. Lett. **97** (2010) 042509 [71]. Copyright 2010 American Institute of Physics.

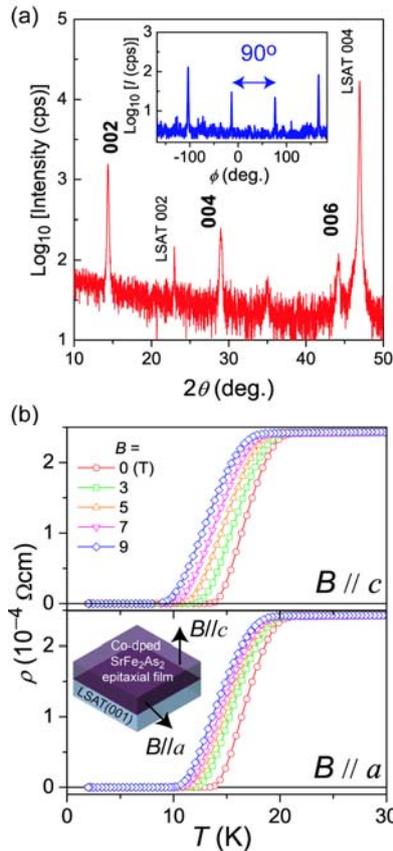

Fig. 6. (Color online) (a) Out-of-plane XRD pattern of Sr122:Co epitaxial film grown at ~700 °C on LSAT (001) substrate with a Nd:YAG laser ablation system. The inset shows an in-plane $\phi$-scan of the Sr122:Co 200 diffraction. (b) $\rho$–$T$ curves of the film and their magnetic anisotropy. Reprinted from H. Hiramatsu et al.: Appl. Phys. Express **1** (2008) 101702 [79]. Copyright 2008 Japan Society of Applied Physics.

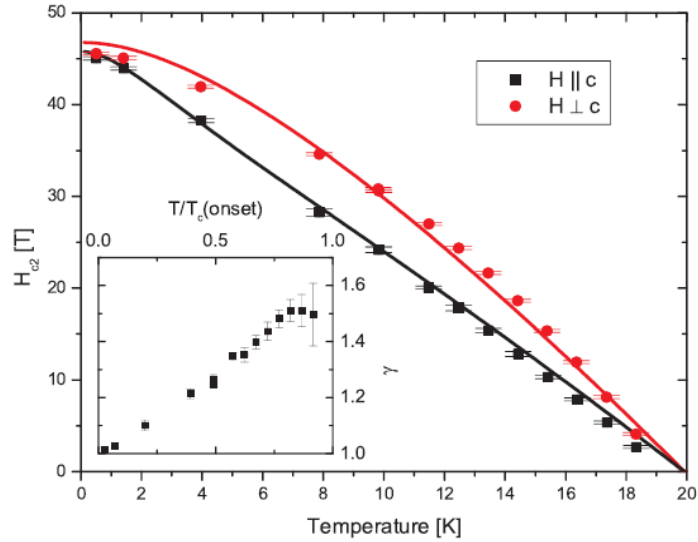

Fig. 7. (Color online) Temperature dependence of upper critical magnetic field of Sr122:Co epitaxial film for $H \parallel c$ (squares) and $H \perp c$ (circles) under strong magnetic fields. The inset shows that the anisotropy ($\gamma$) decreases linearly with decreasing temperature. Reprinted from S. A. Baily et al.: Phys. Rev. Lett. **102** (2009) 117004 [80]. Copyright 2009 American Physical Society.

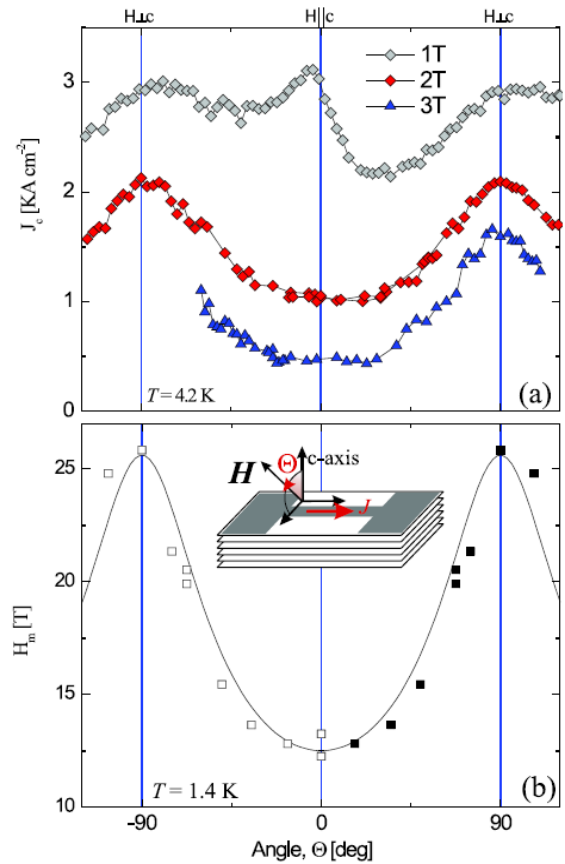

Fig. 8. (Color online) Angular dependence of $J_c$ at 4.2 K (a) and melting line at 1.4 K (b) of Sr122:Co epitaxial film. Reprinted from B. Maiorov et al.: Supercond. Sci. Technol. **22** (2009) 125011 [81]. Copyright 2009 Institute of Physics Publishing.

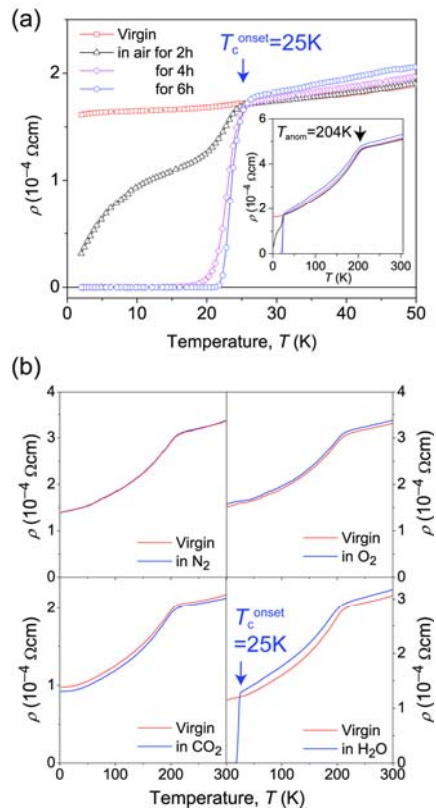

Fig. 9. (Color online) (a) Change in $\rho$–$T$ curves as a function of air-exposure time for undoped Sr122 epitaxial films. The inset shows the curves in the range 2–305 K. (b) Effects of exposing the thin films to the constituents of air individually. Reprinted from H. Hiramatsu et al.: Phys. Rev. B **80** (2009) 052501 [83]. Copyright 2009 American Physical Society.

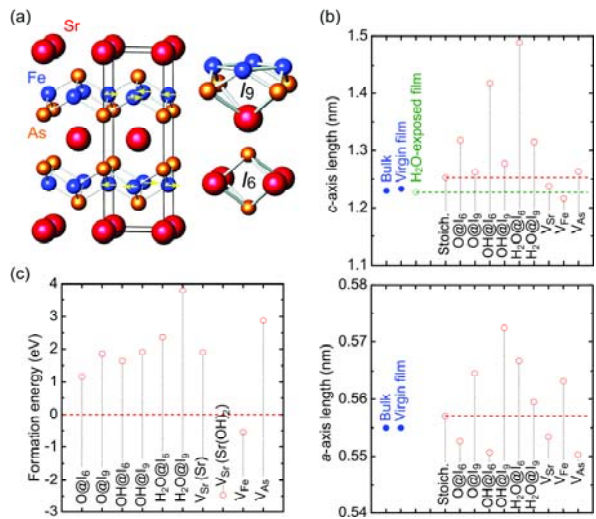

Fig. 10. (Color online) (a) Two interstitial sites ($I_9$ and $I_6$) in Sr122 phase. The box indicates the unit cell. (b) Calculated lattice parameters of impurity- or vacancy-containing Sr122. (c) Formation energies of impurities and vacancies. For $V_{Sr}$, those for two reactions, $SrFe_2As_2 \rightarrow SrFe_2As_2{:}V_{Sr} + Sr$ ('Sr') and $SrFe_2As_2 + 2H_2O \rightarrow SrFe_2As_2{:}V_{Sr} + Sr(OH)_2 + H_2$ [ 'Sr(OH)$_2$'], are shown. Reprinted from T. Kamiya et al.: Mater. Sci. Eng. B **173** (2010) 244 [93]. Copyright 2010 Elsevier.

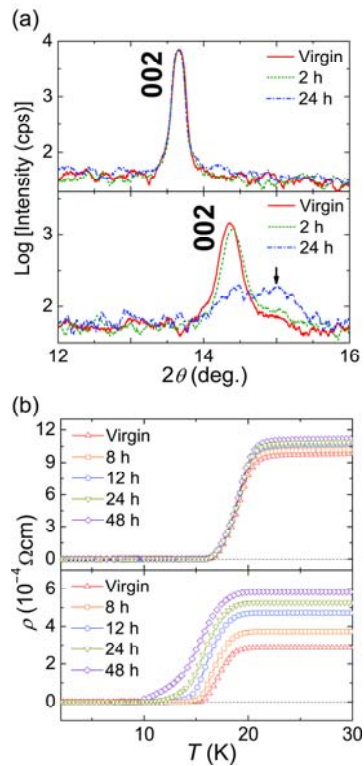

Fig. 11. (Color online) Variation of XRD patterns around the 002 diffraction peak (a) and $\rho$–$T$ curves (b) as a function of water-vapor exposure time for Ba122:Co (top) and Sr122:Co (bottom) epitaxial films. The peak indicated by an arrow in (a) is assigned to Fe$_2$As. Reprinted from T. Katase et al.: Solid State Commun. **149** (2009) 2121 [95]. Copyright 2009 Elsevier.

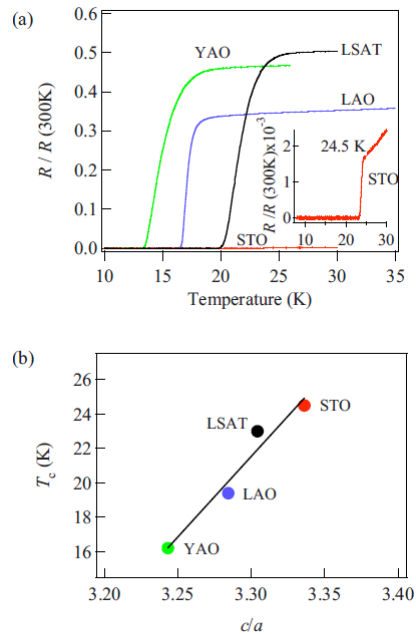

Fig. 12. (Color online) (a) Temperature dependence of normalized resistance of Ba122:Co epitaxial films grown on four types of single-crystal substrates. (b) Relation between $T_c$ and $c/a$ ratio. Reprinted from K. Iida et al.: Appl. Phys. Lett. **95** (2009) 192501 [98]. Copyright 2009 American Institute of Physics.

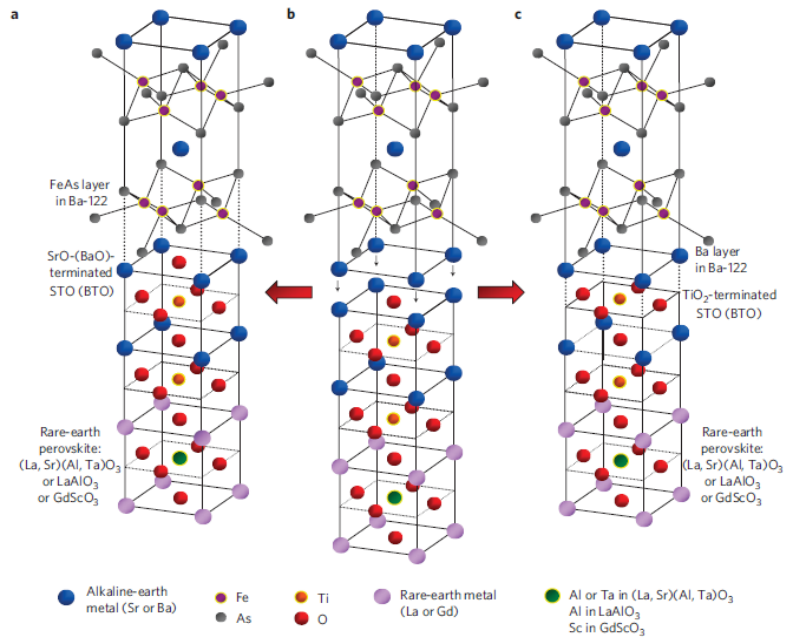

Fig. 13. (Color online) Schematically illustrated concept of using perovskite-oxide buffer to obtain high-quality Ba122:Co epitaxial films on LSAT, LAO, etc. Reprinted from S. Lee et al.: Nat. Mater. **9** (2010) 397 [100]. Copyright 2010 Nature Publishing Group.

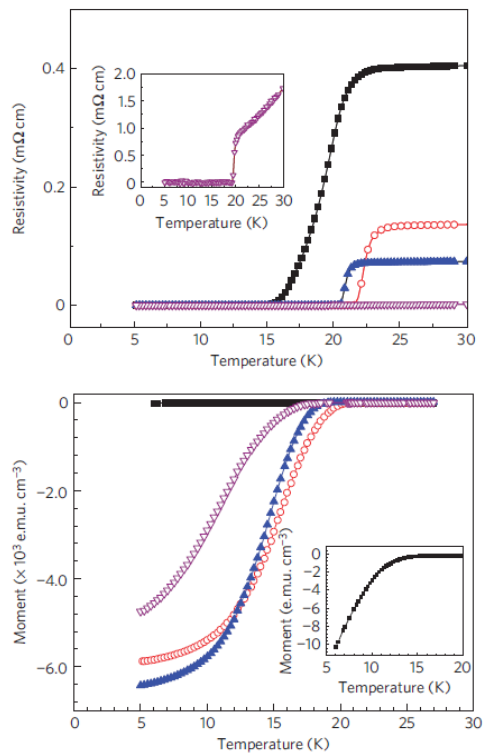

Fig. 14. (Color online) $\rho$–$T$ curves (top) and magnetic properties at 2 mT after zero-field cooling (bottom) of Ba122:Co epitaxial films on bare LSAT (black squares), 100 u.c. STO / LSAT (red circles), 50 u.c. BTO / LSAT (blue closed triangles), and bare STO (purple open triangles). Reprinted from S. Lee et al.: Nat. Mater. **9** (2010) 397 [100]. Copyright 2010 Nature Publishing Group.

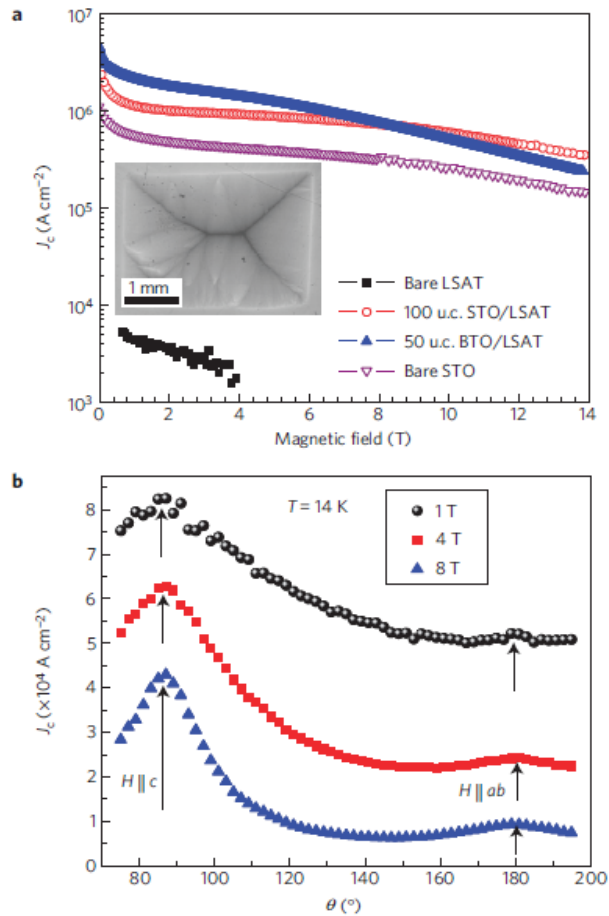

Fig. 15. (Color online) (a) Magnetic $J_c$–$B$ curves of Ba122:Co epitaxial films on bare LSAT (black squares), 100 u.c. STO / LSAT (red circles), 50 u.c. BTO / LSAT (blue closed triangles), and bare STO (purple open triangles) at 4.2 K. The inset shows a magneto-optical image of the film on 100 u.c. STO / LSAT at 6.6 K. (b) Angular dependence of transport $J_c$ of the Ba122:Co epitaxial film on STO / LSAT substrate at 1, 4, and 8 T. Reprinted from S. Lee et al.: Nat. Mater. **9** (2010) 397 [100]. Copyright 2010 Nature Publishing Group.

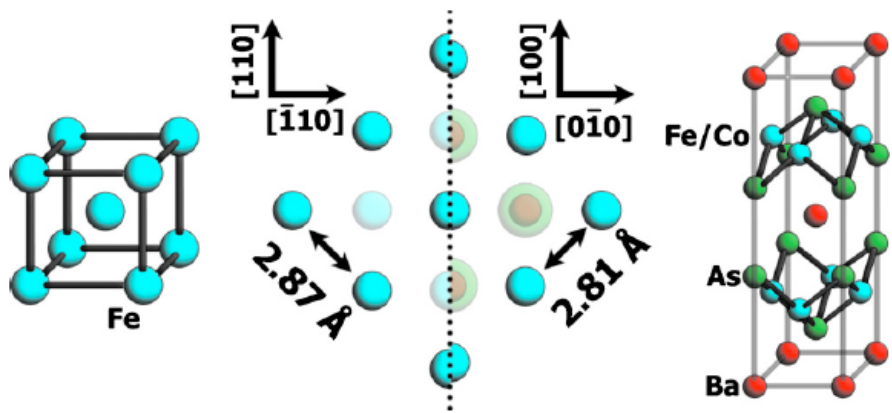

Fig. 16. (Color online) In-plane lattice mismatch between Fe and Ba122. Reprinted from T. Thersleff et al.: Appl. Phys. Lett. **97** (2010) 022506 [105]. Copyright 2010 American Institute of Physics.

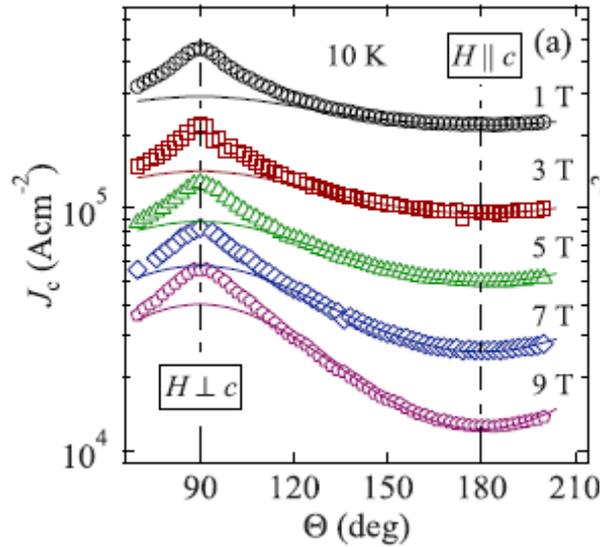

Fig. 17. (Color online) Angular dependence of $J_c$ at 10 K and magnetic fields of 1–9 T for Ba122:Co epitaxial film on an Fe (15 nm in thickness) buffer layer. Reprinted from K. Iida et al.: Appl. Phys. Lett. **97** (2010) 172507 [104]. Copyright 2010 American Institute of Physics.

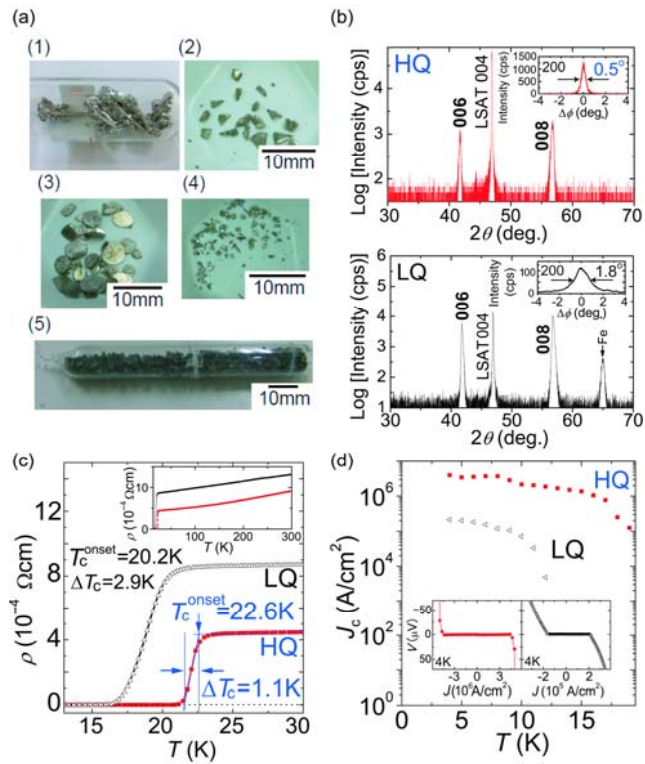

Fig. 18. (Color online) (a) Preparation procedure of BaAs precursor for Ba122:Co laser ablation target. (b) Out-of-plane XRD patterns, (c) $\rho$–$T$ curves, and (d) $J_c$–$T$ curves of Ba122:Co epitaxial films (LQ: black) and the improved films (HQ, red). The insets in (b), (c), and (d) show the in-plane $\phi$-scan of the 200 diffraction, the $\rho$–$T$ curve at $T = 2$–305 K, and the $I$–$V$ curve at 4 K for each film, respectively. Reprinted from T. Katase et al.: Appl. Phys. Express **3** (2010) 063101 [107]. Copyright 2010 Japan Society of Applied Physics.

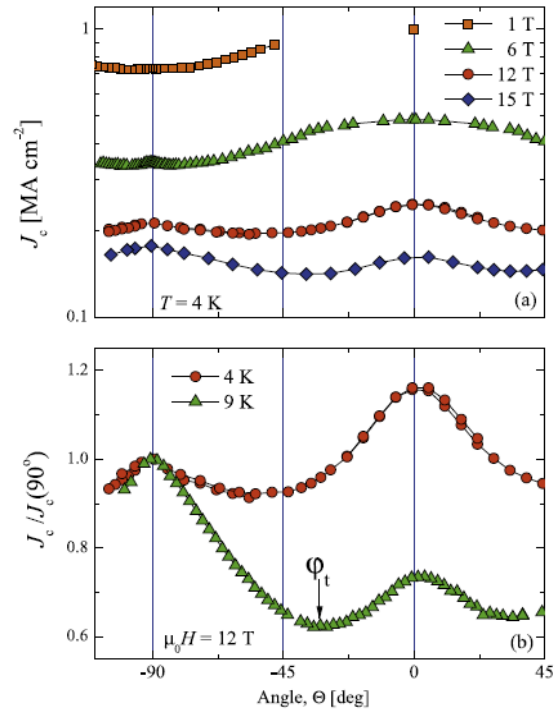

Fig. 19. (Color online) (a) Angular dependence of $J_c$ at 4 K and magnetic fields of 1–15 T for Ba122:Co epitaxial film grown on LSAT (001) substrate. (b) Normalized $J_c(\theta)$ of the film under 12 T at 4 and 9 K. Reprinted from B. Maiorov et al.: Supercond. Sci. Technol. **24** (2011) 055007 [111]. Copyright 2011 Institute of Physics Publishing.

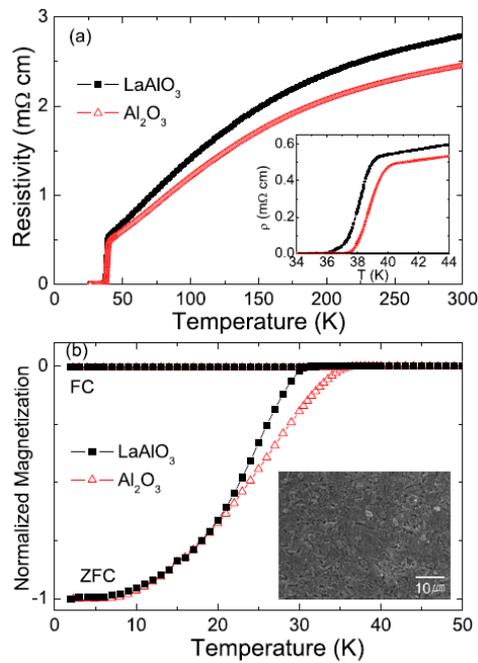

Fig. 20. (Color online) Temperature dependence of resistivity (a) and magnetization (b) of Ba122:K thin films grown by *ex situ* laser ablation on LAO (001) and α-$Al_2O_3$ (0001) single-crystal substrates. The inset of (b) shows a scanning electron microscopy image of the film surface. Reprinted from N. H. Lee et al.: Appl. Phys. Lett. **96** (2010) 202505 [112]. Copyright 2010 American Institute of Physics.

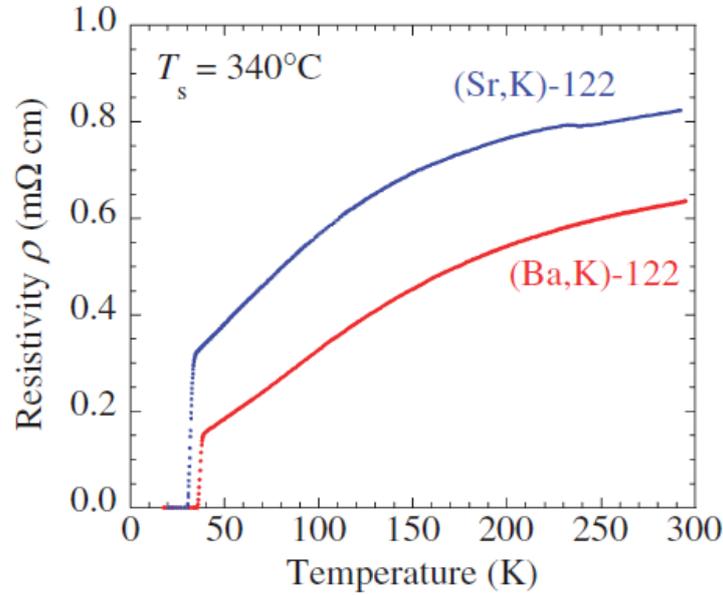

Fig. 21. (Color online) Temperature dependence of resistivity of Ba122:K and Sr122:K epitaxial films grown by MBE at 340 °C using reduced As flux on r-sapphire single-crystal substrates. Reprinted from S. Takeda et al.: Appl. Phys. Express **3** (2010) 093101 [113]. Copyright 2010 Japan Society of Applied Physics.

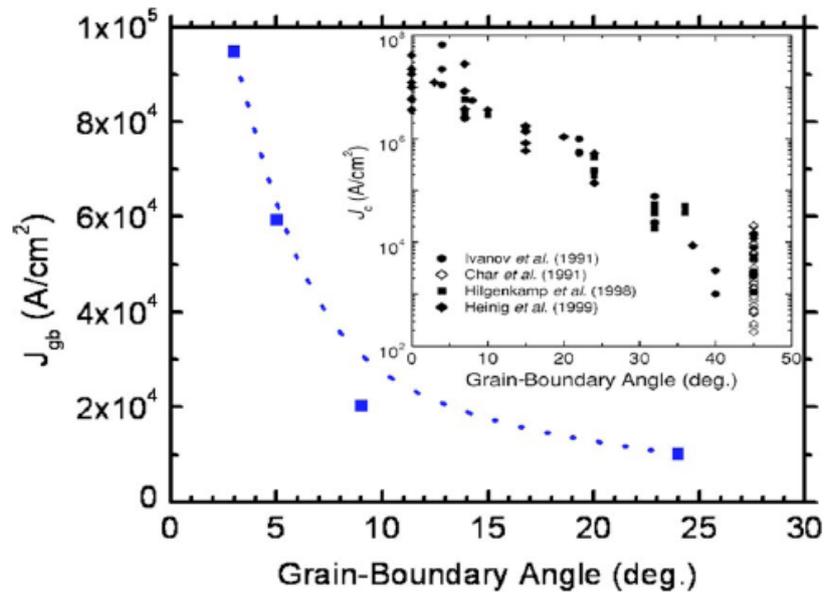

Fig. 22. (Color online) (a) Misorientation angle of GB ($\theta_{GB}$) dependence of $J_c$ of Ba122:Co on [001]-tilt STO bicrystal substrates at 12 K under an external magnetic field of 0.5 T. The inset shows the corresponding results for YBCO. Reprinted from S. Lee et al.: Appl. Phys. Lett. **95** (2009) 212505 [97]. Copyright 2009 American Institute of Physics.

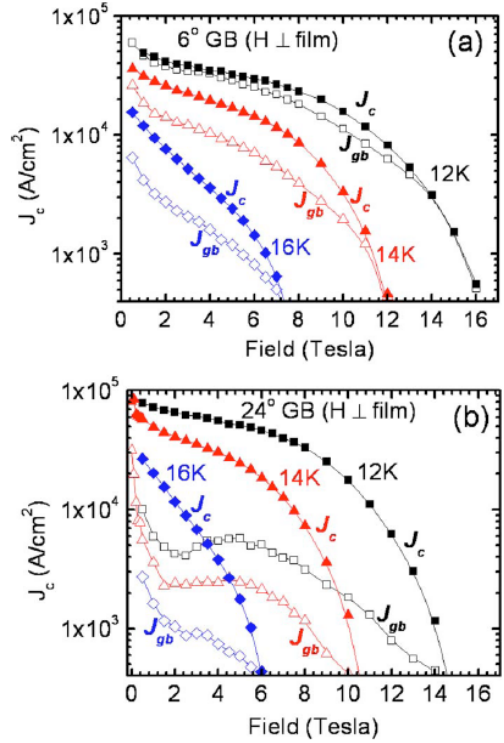

Fig. 23. (Color online) Magnetic field dependence of $J_{gb}$ ($J_c$ with BGB) and $J_c$ ($J_c$ without BGB) of Ba122:Co on (a) $\theta_{GB} = 6°$ STO bicrystal and (b) $\theta_{GB} = 24°$ STO bicrystal at 12, 14, and 16 K. Reprinted from S. Lee et al.: Appl. Phys. Lett. **95** (2009) 212505 [97]. Copyright 2009 American Institute of Physics.

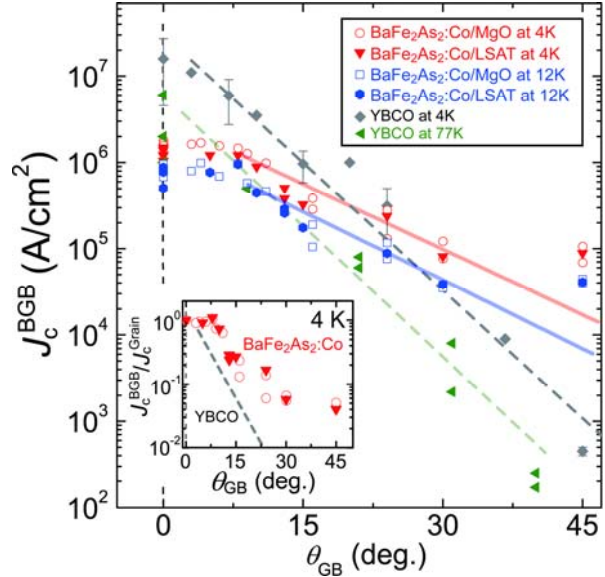

Fig. 24. (Color online) Misorientation angle of GB ($\theta_{GB}$) dependence of self-field intergrain $J_c$ ($J_c^{BGB}$) in Ba122:Co BGB junctions grown on [001]-tilt bicrystal substrates of MgO and LSAT. The red and blue solid lines are fitted using the empirical equation $J_c^{BGB} = J_{c0}\exp(-\theta_{GB}/\theta_0)$ in the weak-link regime. Typical data (gray: at 4 K, green: at 77 K) for YBCO BGB junctions are also shown for comparison. The inset shows the ratio of $J_c^{BGB}$ to the intragrain $J_c$ ($J_c^{Grain}$) at 4 K. Reprinted from T. Katase et al.: Nat. Commun. **2** (2011) 409 [99]. Copyright 2011 Nature Publishing Group.

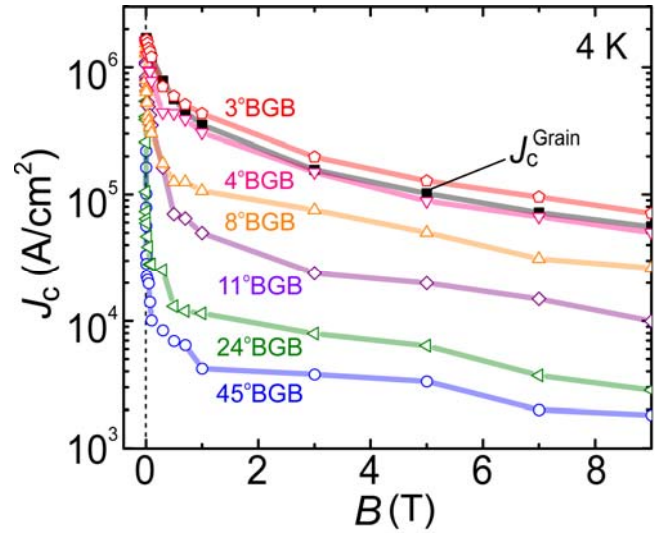

Fig. 25. (Color online) $J_c^{BGB}$–$B$ curves at 4 K for Ba122:Co BGB junctions with $\theta_{GB}$ = 3–45° grown on MgO bicrystal substrates. $J_c^{Grain}$ values measured in a grain bridge on a 3° MgO bicrystal substrate are also plotted as black squares. Reprinted from T. Katase et al.: Nat. Commun. **2** (2011) 409 [99]. Copyright 2011 Nature Publishing Group.

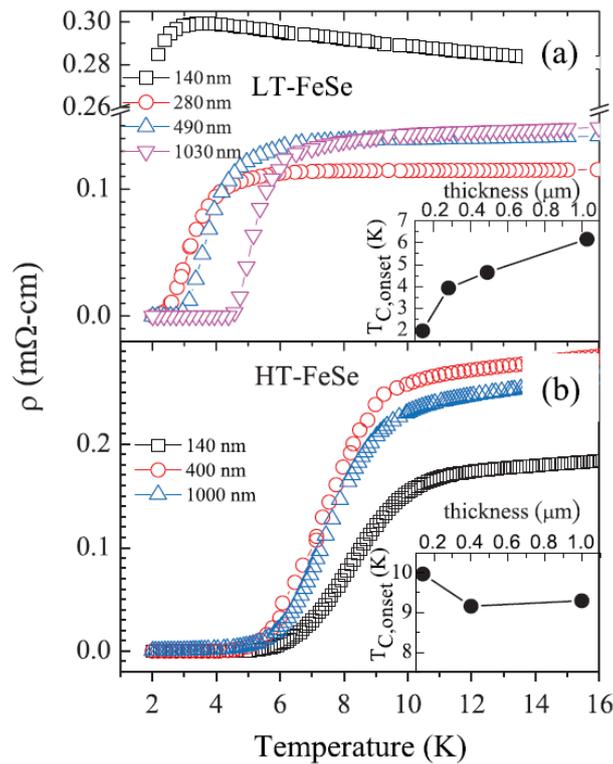

Fig. 26. (Color online) Temperature dependence of resistivity of FeSe$_{1-x}$ films on MgO (001) substrates grown at (a) 320 °C [LT, (001) orientation] and (b) 500 °C [HT, (101) orientation] with different thicknesses between 140 and 1030 nm. The insets show the changes in $T_c^{onset}$ with the thickness. Reprinted from M. J. Wang et al.: Phys. Rev. Lett. **103** (2009) 117002 [145]. Copyright 2009 American Physical Society.

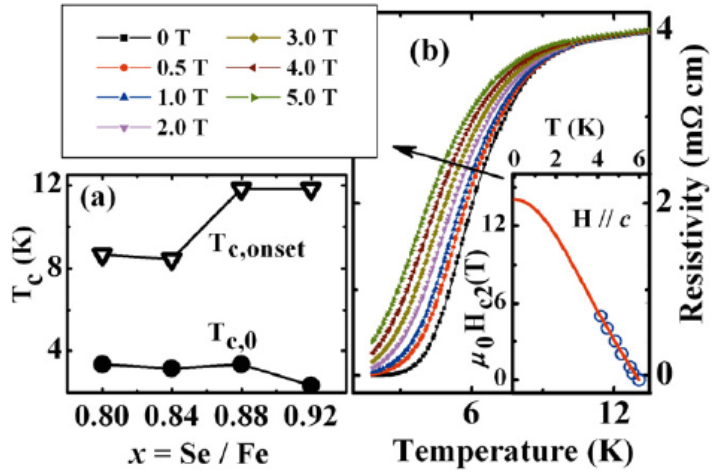

Fig. 27. (Color online) (a) Se content dependence of $T_c^{onset}$ and $T_c^{zero}$ of FeSe$_x$ ($x = 0.80 – 0.92$) films of ca. 200 nm thickness grown at 620 °C on LAO. (b) Magnetic field dependence of the film with $x = 0.88$. The inset shows that the upper critical magnetic field was estimated to be ~14 T. Reprinted from Y. Han et al.: J. Phys.: Condens. Matter **21** (2009) 235702. [146]. Copyright 2009 Institute of Physics Publishing.

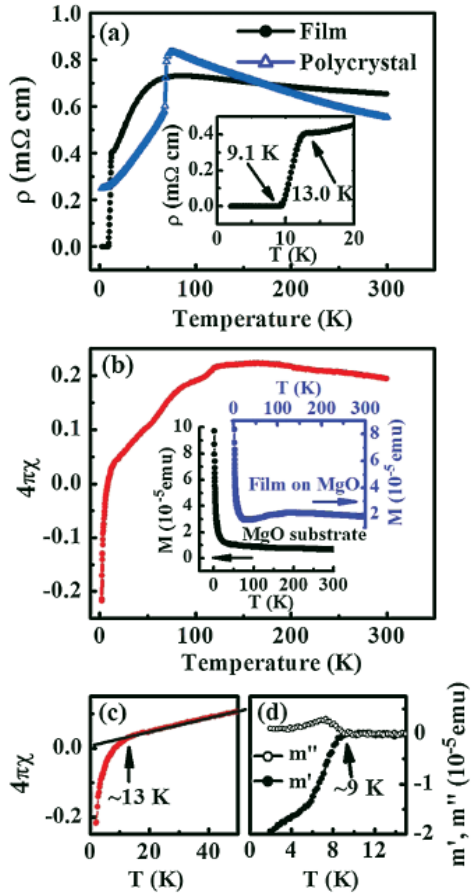

Fig. 28. (Color online) (a) $\rho$–$T$ curve of an FeTe film on MgO. That of a polycrystalline bulk sample is shown for comparison. (b) Temperature dependence of dc magnetic susceptibility of the film. The shielding volume fraction of the film was estimated to be 22% after subtracting the signal from the MgO substrate (Inset). (c) Enlarged image of (b). (d) ac susceptibility of the film. Reprinted from Y. Han et al.: Phys. Rev. Lett. **104** (2010) 017003 [155]. Copyright 2010 American Physical Society.

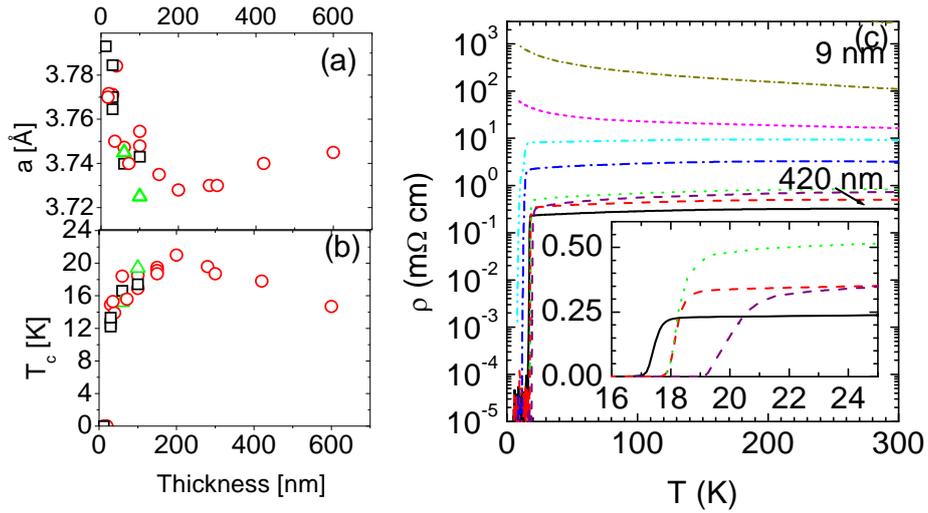

Fig. 29. (Color online) Film thickness dependence of (a) $a$-axis length and (b) $T_c^{onset}$ of Fe(Se,Te) epitaxial films on LAO (circles), STO (squares), and YSZ (triangles) single-crystal substrates. (c) $\rho$–$T$ curves of Fe(Se,Te) epitaxial films on LAO with thicknesses of 9–420 nm. Reprinted from E. Bellingeri et al.: Appl. Phys. Lett. **96** (2010) 102512 [158]. Copyright 2010 American Institute of Physics.

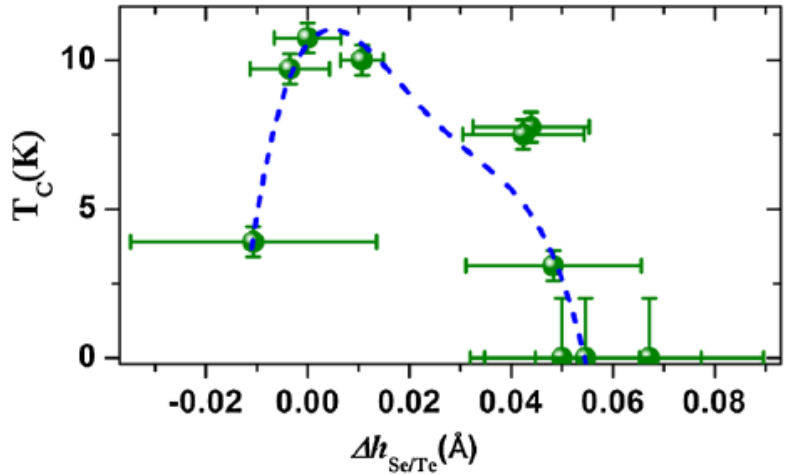

Fig. 30. (Color online) $T_c$ versus $\Delta h_{Se/Te}$ (= $c\Delta z + z\Delta c$) for FeSe$_{0.5}$Te$_{0.5}$ films grown on MgO at different temperatures. Reprinted from S. X. Huang et al.: Phys. Rev. Lett. **104** (2010) 217002 [161]. Copyright 2010 American Physical Society.

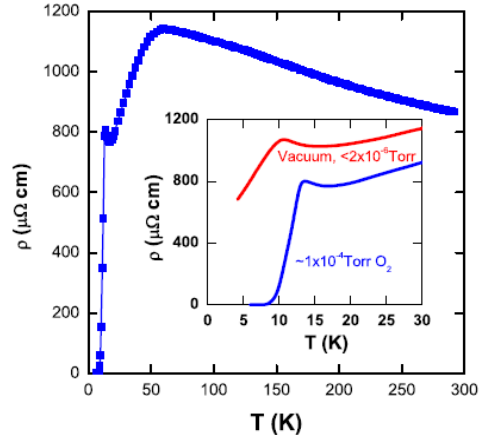

Fig. 31. (Color online) $\rho$–$T$ curve of $Fe_{1.08}Te:O_x$ epitaxial film on STO substrate. The inset shows $\rho$–$T$ curves for two types of $Fe_{1.08}Te$ films grown in oxygen (~$1 \times 10^{-4}$ Torr) and vacuum (< $2 \times 10^{-6}$ Torr) atmospheres. Reprinted from W. Si et al.: Phys. Rev. B **81** (2010) 092506 [156]. Copyright 2010 American Physical Society.

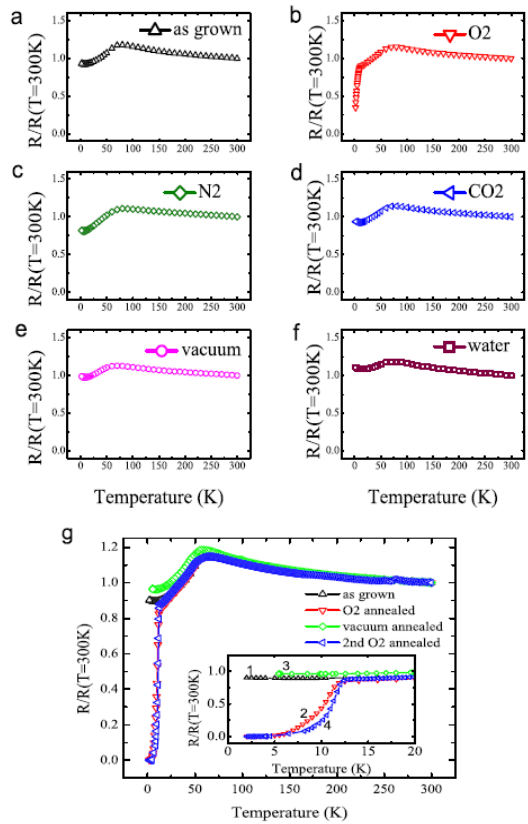

Fig. 32. (Color online) $\rho$–$T$ curves of FeTe films on MgO after treatment under different atmospheres. (a) As-grown, and annealed in (b) $O_2$, (c) $N_2$, (d) $CO_2$ ($1 \times 10^{-1}$ Torr at 100 °C), and (e) in vacuum ($2 \times 10^{-7}$ Torr at 100 °C), and (f) exposure to water (40 °C). (g) Change in the $\rho$–$T$ curves after repeated oxygen and vacuum annealing treatments. Reprinted from Y. F. Nie et al.: Phys. Rev. B **82** (2010) 020508 [157]. Copyright 2010 American Physical Society.

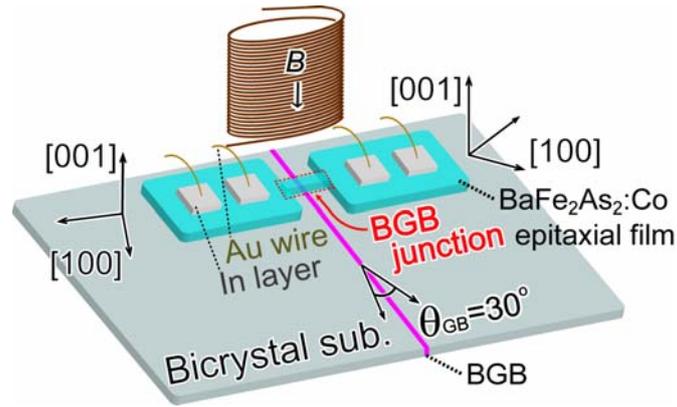

Fig. 33. (Color online) Device structure of 10-μm-wide Ba122:Co BGB junction fabricated on [001]-tilt LSAT bicrystal substrates with $\theta_{GB} = 30^o$ [108].

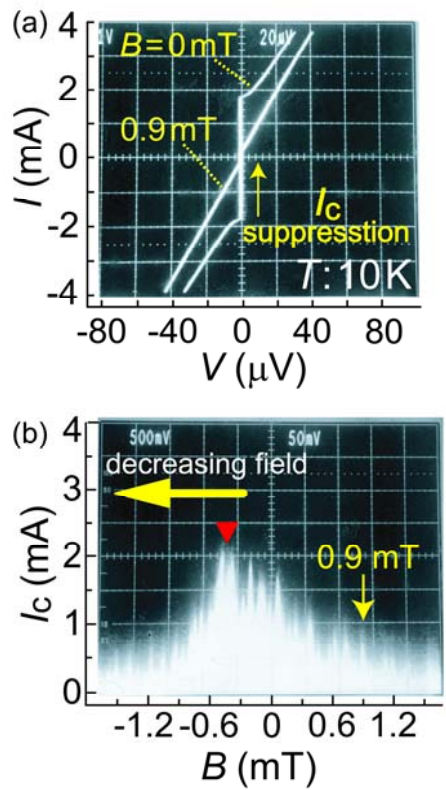

Fig. 34. (Color online) (a) I–V characteristic of Ba122:Co BGB junction with $\theta_{GB} = 30^o$ under external magnetic fields $B = 0$ and 0.9 mT at 10 K. (b) Magnetic field dependence of critical current ($I_c$) at 10 K. The magnetic field sweep direction and the position of maximum $I_c$ are indicated by the horizontal yellow arrow and red triangle, respectively. Reprinted from T. Katase et al.: Appl. Phys. Lett. **96** (2010) 142507 [108]. Copyright 2010 American Institute of Physics.

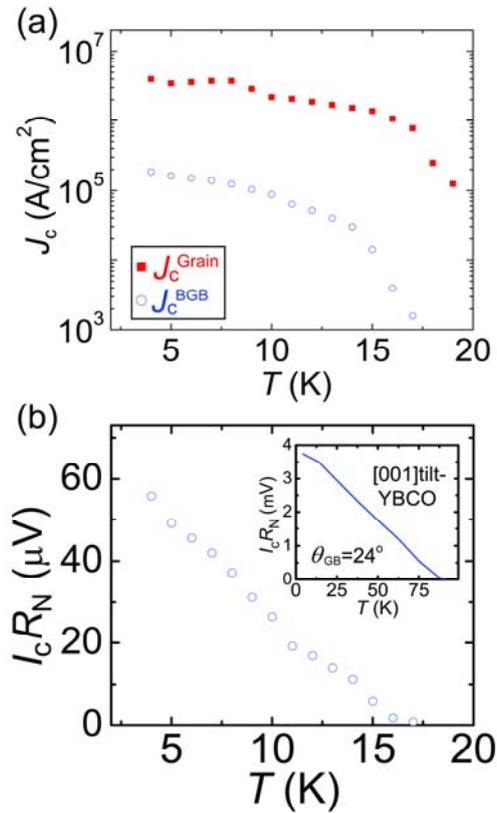

Fig. 35. (Color online) (a) Temperature dependence of self-field $J_c$ of grain (non-BGB, $J_c^{Grain}$) and BGB bridges ($J_c^{BGB}$) of Ba122:Co film on LSAT bicrystal substrate with $\theta_{GB} = 30°$. (b) Temperature dependence of $I_cR_N$ of Ba122:Co BGB junction along with that of YBCO BGB junction with $\theta_{GB} = 24°$. Reprinted from T. Katase et al.: Appl. Phys. Lett. **96** (2010) 142507 [108]. Copyright 2010 American Institute of Physics.

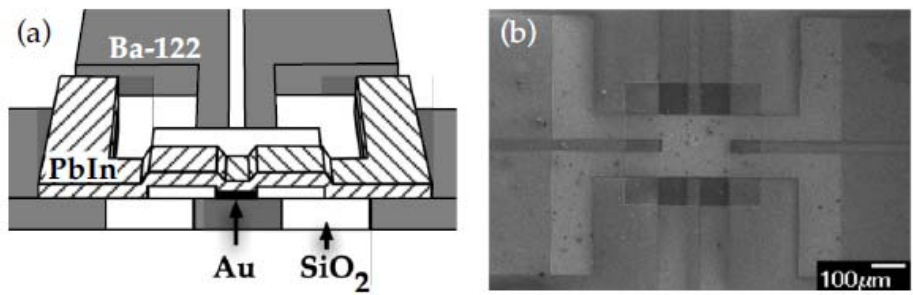

Fig. 36. (a) Device structure of thin film hybrid SNS junction with Ba122:Co/Au/PbIn structure. (b) Scanning electron microscopic top image of the device. Reprinted from S. Schmidt et al.: Appl. Phys. Lett. **97** (2010) 172504 [189]. Copyright 2010 American Institute of Physics.

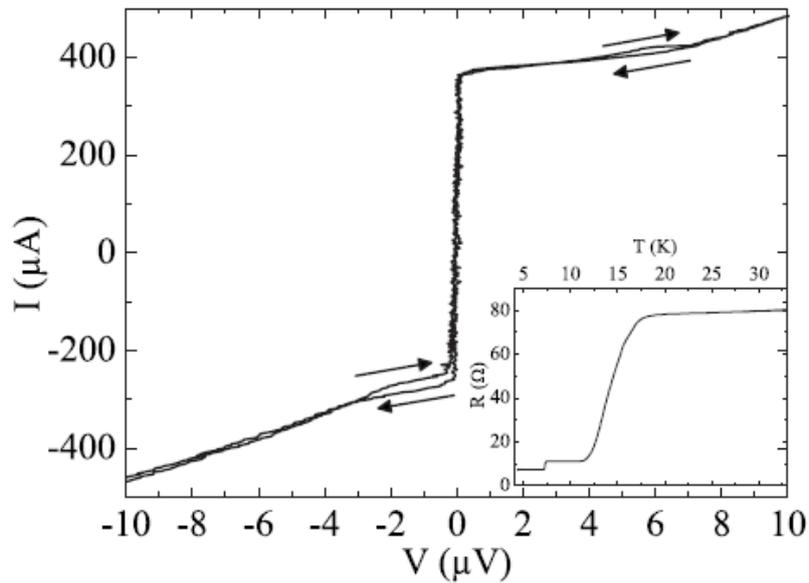

Fig. 37. *I–V* characteristic of Ba122:Co/Au/PbIn SNS junction measured bidirectionally at 4.2 K. The inset shows the temperature dependence of resistance. Reprinted from S. Schmidt et al.: Appl. Phys. Lett. **97** (2010) 172504 [189]. Copyright 2010 American Institute of Physics.

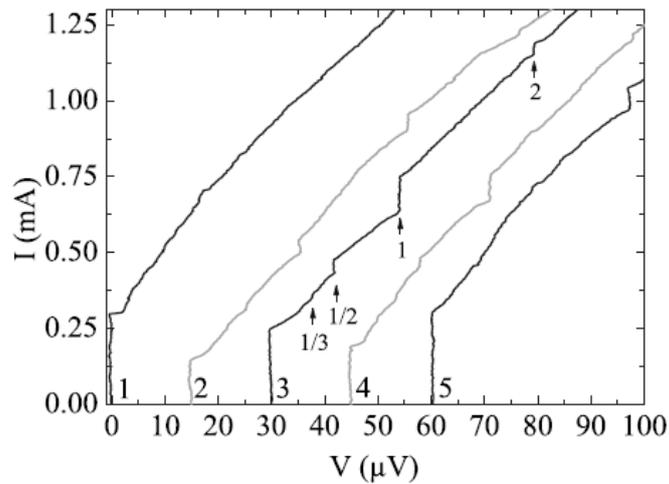

Fig. 38. *I–V* curves of Ba122:Co/Au/PbIn SNS junction under microwave irradiation at various frequencies (1: without microwave, 2: 10 GHz, 3: 12 GHz, 4: 13 GHz, and 5: 18 GHz). Reprinted from S. Schmidt et al.: Appl. Phys. Lett. **97** (2010) 172504 [189]. Copyright 2010 American Institute of Physics.

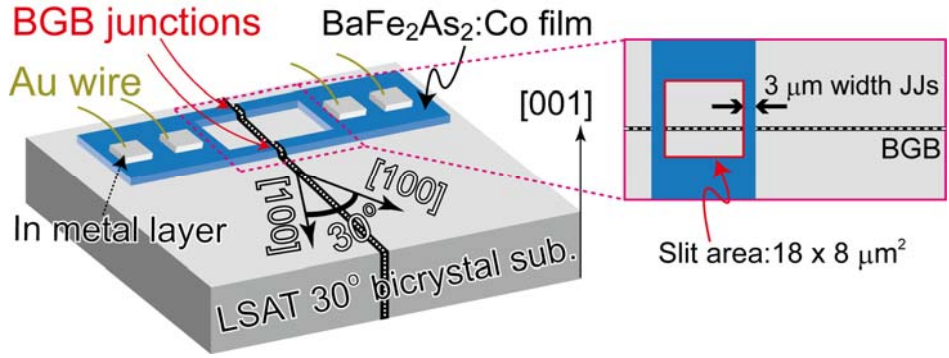

Fig. 39. (Color online) Schematic image of dc-SQUID structure consisting of two 3-μm-wide Josephson junctions fabricated on [001]-tilt LSAT bicrystal substrate with $\theta_{GB} = 30°$. A Ba122:Co superconducting loop with a slit area of $18 \times 8$ μm$^2$ was located across the BGB. Reprinted from T. Katase et al.: Supercond. Sci. Technol. **23** (2010) 082001 [109]. Copyright 2010 Institute of Physics Publishing.

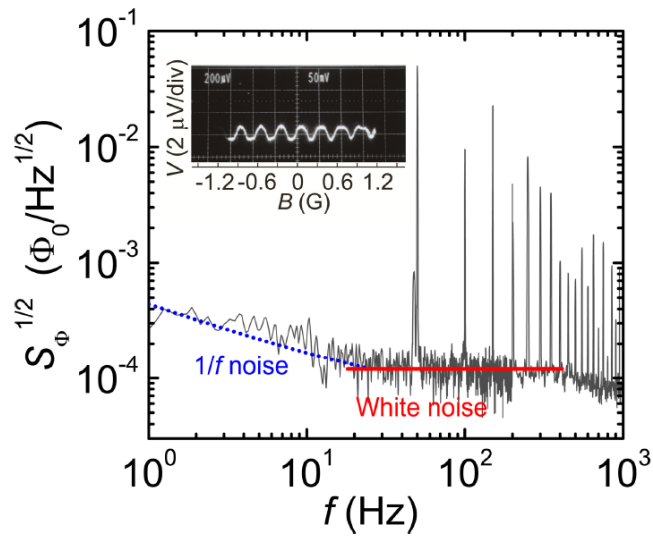

Fig. 40. (Color online) Flux noise ($S_\Phi^{1/2}$) of Ba122:Co dc-SQUID as a function of frequency ($f$) measured using dc bias mode of an FLL circuit at 14 K. The white noise region ($f > 20$ Hz) and $1/f$ noise region ($f < 20$ Hz) are indicated by solid and dotted lines, respectively. The inset shows the voltage–flux ($V$–$\Phi$) characteristics at 14 K. Reprinted from T. Katase et al.: Supercond. Sci. Technol. **23** (2010) 082001 [109]. Copyright 2010 Institute of Physics Publishing.

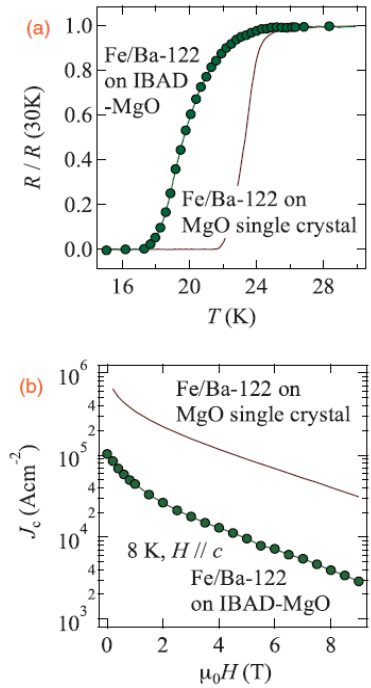

Fig. 41. (Color online) (a) Temperature dependence of normalized resistivity and (b) magnetic field dependence of $J_c$ at 8 K for Ba122:Co thin film grown on Fe buffer / IBAD-MgO tape substrate. Results in the case of an Fe buffer / MgO single-crystal substrate are also shown for comparison. Reprinted from K. Iida et al.: Appl. Phys. Express **4** (2011) 013103 [205]. Copyright 2011 Japan Society of Applied Physics.

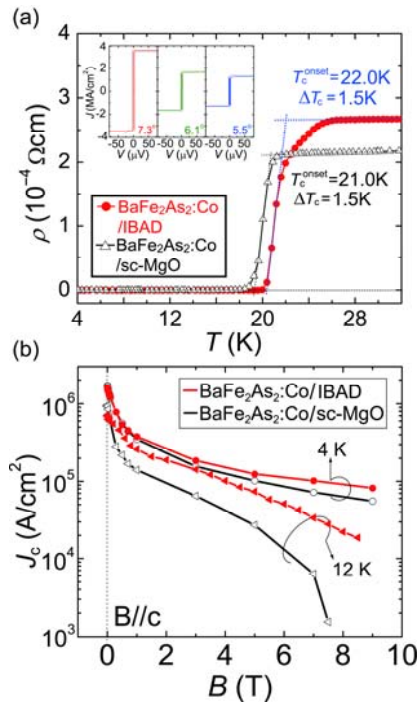

Fig. 42. (Color online) (a) $\rho$–$T$ curves and (b) $J_c(B)$ of Ba122:Co films on IBAD-MgO tape substrate. Results in the case of a MgO single-crystal substrate are shown for comparison. The inset in (a) shows the $I$–$V$ curves at 2 K for the Ba122:Co films on IBAD-MgO tape substrates with different $\Delta\phi_{MgO}$ of 7.3, 6.1, and 5.5°. Reprinted from T. Katase et al.: Appl. Phys. Lett. **98** (2011) 242510 [206]. Copyright 2011 American Institute of Physics.